\documentclass[aps,preprint,prd,showpacs,showkeys,nofootinbib,superscriptaddress,tightenlines]{revtex4-1}
\usepackage{amsmath, amssymb, amsfonts}
\usepackage{graphicx}
\usepackage{color}
\usepackage{xcolor}
\usepackage{lineno}
\usepackage{multirow}
\usepackage{bm}
\usepackage{ulem}

\normalsize
\usepackage{booktabs}

\newcommand{\bea}{\begin{eqnarray}}
\newcommand{\eea}{\end{eqnarray}}
\newcommand{\beq}{\begin{equation}}
\newcommand{\eeq}{\end{equation}}
\newcommand{\bqa}{\begin{eqnarray}}
\newcommand{\eqa}{\end{eqnarray}}

\graphicspath{{Figures/}}

\begin{document}
\title{Thermal Energy of a Charm-meson Molecule \\in a Pion Gas}
\author{Eric Braaten}
\email{braaten.1@osu.edu}
\affiliation{Department of Physics,
	         The Ohio State University, Columbus, OH\ 43210, USA}
	
\author{Li-Ping He}
\email{heliping@hiskp.uni-bonn.de}
\affiliation{Helmholtz-Institut f\"ur Strahlen- und Kernphysik and Bethe Center for Theoretical Physics, Universit\"at Bonn, D-53115 Bonn, Germany}
	
\author{Kevin Ingles}
\email{ingles.27@buckeyemail.osu.edu}
\affiliation{Department of Physics,
	         The Ohio State University, Columbus, OH\ 43210, USA}
	
\author{Jun Jiang}
\email{jiangjun87@sdu.edu.cn}
\affiliation{School of Physics, Shandong University, Jinan, Shandong 250100, China}
\date{\today}

\begin{abstract}
The thermal corrections to the propagator of a loosely bound charm-meson molecule in a pion gas
are calculated to next-to-leading order in the heavy-meson expansion using a zero-range effective field theory.
Ultraviolet divergences in the charm-meson-pair self energy 
are canceled by corrections to the charm-meson-pair contact vertex.
Terms that are singular at the charm-meson-pair threshold  can be absorbed 
into thermal corrections to the rest energies and kinetic masses of the charm-meson constituents.
The remaining terms reduce to a thermal correction to the binding momentum that is
proportional to the pion number density and suppressed by the pion/charm-meson mass ratio.
The correction gives a tiny decrease in the binding energy of the charm-meson molecule 
relative to the charm-meson-pair threshold in the pion gas
and a change in its thermal width that is small compared to the thermal widths of the charm-meson constituents.
These results are encouraging for the prospects of observing $X(3872)$ and $T_{cc}^+(3875)$
in the expanding hadron gas produced by heavy-ion collisions.
\end{abstract}


\maketitle
	
\section{Introduction}

Nature has provided us with at least two exquisite examples of loosely bound hadronic molecules:
$X(3872)$ (or, more concisely, $X$), also known as  $\chi_{c1}(3872)$,
which was discovered by the Belle collaboration in 2003 \cite{Belle:2003nnu},
and $T_{cc}^+(3875)$ (or, more concisely, $T_{cc}^+$), which was discovered by the LHCb collaboration in 2021 \cite{LHCb:2021vvq}.
Since their discovery, there have been numerous studies of these and other exotic heavy hadrons.
For recent reviews, see Refs.~\cite{Brambilla:2019esw,Chen:2022asf}.
Their binding energies relative to the appropriate charm-meson-pair thresholds
and their decay widths are both over an order of magnitude smaller than the energy scale 
associated with pion exchange between the charm mesons: $m_\pi^2/M \approx 10$\,MeV,
where $m_\pi$ is the pion mass and $M$ is the charm-meson mass.
In addition, the charm-meson constituents have decay widths  that are over two orders of magnitude smaller 
than $m_\pi^2/M$.
The $X$ and $T_{cc}^+$ are much more loosely bound than the deuteron,
whose binding energy is about 40\% of the energy scale  5\,MeV 
associated with pion exchange between the nucleons.
We refer Ref.~\cite{Braaten:2004rn} for a review of the universality of a loosely bound molecule and Ref.~\cite{Guo:2017jvc} for a general introduction to hadronic molecules.

The tiny binding energies of $X$ and $T_{cc}^+$ can be exploited 
to develop quantitative treatments of their properties, including their interactions with other hadrons.
The simplest effective field theory (EFT) that can be applied to a loosely bound charm-meson molecule
is a zero-range effective field theory (ZREFT) for nonrelativistic charm mesons,
in which they interact only through contact interactions \cite{Braaten:2003he}.
This ZREFT is the analog for charm mesons 
of the pionless EFT that has been applied to the few-nucleon problem \cite{Kaplan:1998tg}.
An alternative EFT with a greater range of validity is XEFT,
which describes nonrelativistic charm mesons and pions \cite{Fleming:2007rp}.
Calculations in XEFT can be simplified by using a Galilean-invariant formulation of XEFT
that exploits the fact that the $D^\ast$-$D$ mass difference is approximately equal to the pion mass 
\cite{Braaten:2015tga,Braaten:2020nmc}.

A challenging problem to which EFT's for charm mesons may be able to provide some insight
is the production of $X$ and $T_{cc}^+$ in relativistic heavy-ion collisions. 
A  central relativistic heavy-ion collision is believed to produce a region of quark-gluon plasma
that remains in local thermal equilibrium as it expands and cools.
At a temperature of 156\,MeV, there is a hadronization transition 
from the quark-gluon plasma to a hadron resonance gas,
which continues to expand and cool in local thermal equilibrium.
At a temperature around 115\,MeV, the hadron gas reaches  kinetic freeze-out
and goes out of thermal equilibrium.
The system then continues to expand by free-streaming of the hadrons.

The CMS collaboration has observed the production of $X(3872)$ in Pb-Pb collisions at the LHC \cite{CMS:2021znk}.
The ratio of the production rates of $X$ and $\psi(2S)$ 
seems to be about an order of magnitude larger than that from hadronic production mechanisms, 
such as prompt production in $pp$ collisions and exclusive production in $B$-meson decays.
It is particularly surprising that a system with a binding energy less than 1\,MeV 
can survive in a hadronic environment whose temperature is greater than 100\,MeV.

The behavior of a loosely bound charm-meson molecule in the hadron resonance gas 
near the hadronization temperature is a very challenging problem.
One complication is the many strongly interacting light-hadron resonances that must be taken into account.
Another complication is the restoration of chiral symmetry near the hadronization temperature,
which requires scalar and pseudoscalar charm mesons to become degenerate 
and  requires axial-vector and vector charm mesons to become degenerate. 

A much simpler problem is the behavior of a loosely bound charm-meson molecule in the hadron gas 
near the kinetic freeze-out temperature.
The most abundant hadrons in the hadron gas are pions.
Near kinetic freeze-out, the abundance of kaons is smaller by about a factor of 5 
and other hadrons are even less abundant.
The hadron gas can therefore be approximated by a pion gas.
The temperature of the pion gas is high enough that the pions must be treated as relativistic particles.
However the temperature is low enough that 
it may be possible to describe pion interactions using a chiral effective field theory $(\chi$EFT).

Since the kinetic  freezeout temperature is orders of magnitude larger than the binding energy 
of a loosely bound molecule, one might expect that ZREFT or XEFT are simply not applicable at such a high temperature.
 We show in this paper that ZREFT can be applied to the loosely bound charm-meson molecule in the pion gas
 by first integrating out thermal pions in favor of temperature-dependent modifications of the parameters of  ZREFT.
 The thermal properties of the molecule can then be calculated in terms of those $T$-dependent parameters.
 We calculate thermal corrections to the parameters of  ZREFT using the heavy-meson expansion,
 which is an expansion in powers of charm-meson kinetic energies divided by the pion mass.
 At leading order (LO) in the heavy-meson expansion, 
 the only $T$-dependent changes in the ZREFT parameters 
are thermal corrections to the complex rest energies of the charm mesons.
At next-to-leading order (NLO) in the heavy-meson expansion,
the changes in the ZREFT parameters  are much more complicated.
However, we find that the only correction to the energy of the loosely bound molecule 
relative to the $T$-dependent charm-meson-pair threshold
comes from a small $T$-dependent correction to the complex binding momentum of the molecule. 

The rest of this paper is organized as follows.
In Section~\ref{sec:Molecule}, we describe the amplitude whose pole corresponds to a  loosely bound molecule. 
In Section~\ref{sec:Charm/Pi}, 
we introduce notation for the properties of charm mesons and pions and we describe
the pion gas that can be produced by a relativistic heavy-ion collision.
In Section~\ref{sec:CharmMesons}, we calculate the self energies of charm mesons in the pion gas
to NLO and we determine thermal corrections to the parameters of ZREFT.
In Section~\ref{sec:CharmPair}, we calculate the self-energy of a charm-meson pair in the pion gas  to NLO.
In Section~\ref{sec:Compare}, we determine the thermal mass shift 
and the thermal width of the  loosely bound charm-meson molecule.
We summarize our results in Section~\ref{sec:Summary}.
In Appendices~\ref{app:PionIntegral} and \ref{app:MomIntegral},
we give results for integrals over the pion momentum distribution and
 for integrals over the relative momentum of a charm-meson pair.
In Appendix~\ref{app:Feynman}, we give the Feynman rules used in our calculations.


\section{Loosely bound molecule}
\label{sec:Molecule}

In this Section, we present the simplest approximation to the amplitude 
whose pole corresponds to a loosely bound molecule.
We also calculate the effects on the amplitude from 
corrections to the propagators of the constituents of the molecule.

\subsection{Pair Propagator}
\label{subsec:PairProp}

\begin{figure}[t]
\includegraphics[width=0.3\textwidth]{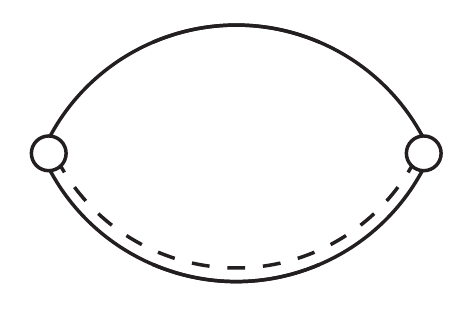}
\caption{
Bubble diagram for the amplitude for  the propagation of $D$ and $D^\ast$ between contact interactions.
The propagator for the $D$ and $D^\ast$ are represented by a solid line and a double (solid+dashed) line, respectively.
The Feynman rule for the open circles is 1.
}
\label{fig:DstarDbubble}
\end{figure}

We consider a loosely bound molecule $X$ whose two constituents are  
a particle $D$ with kinetic mass $M$ and rest energy $\varepsilon$
and a particle $D^\ast$ with kinetic mass $M_\ast$ and rest energy $\varepsilon_\ast$.
The nonrelativistic  propagator for $D$ with energy $E$ (relative to $M$) and momentum $\bm{p}$ is
$i/[E - \varepsilon - \bm{p}^2/(2M) + i \epsilon]$.
The propagator for $D^\ast$ has an analogous expression.
The amplitude for the propagation of $D$ and $D^*$ between contact interactions 
can be represented by the bubble diagram in Fig.~\ref{fig:DstarDbubble}.
The amplitude with nonrelativistic  propagators is linearly ultraviolet divergent.
The ultraviolet cutoff on the loop momentum $\bm{k}$
is most conveniently implemented by subtracting and adding to the integrand 
a term proportional to  $1/k^2$ and then imposing the cutoff on the additional term.
The amplitude for total energy $E$ (relative to $M+M_\ast$) and  total momentum $\bm{P}$ is
\bqa
&&i \bigg[ \int \frac{d^3k}{(2\pi)^3}
\left( \frac{1}{E  - (\varepsilon_\ast +\varepsilon) - \bm{k}^2/(2M) - (\bm{P}-\bm{k})^2/(2M_\ast) + i \epsilon} 
+\frac{2\mu}{k^2} \right)
\nonumber\\
&&\hspace{1cm}
-2 \mu  \int \frac{d^3k}{(2\pi)^3} \frac{1}{k^2} \bigg] = 
- i \frac{\mu}{2\pi}\, \big[ \Lambda - S_0(E_\mathrm{cm} ) \big],
\label{bubble}
\eqa
where $\mu = M_\ast M /(M_\ast  + M)$ is the $D^\ast D$ reduced mass.
We have imposed a momentum cutoff $|\bm{k}| < (\pi/2) \Lambda$ on the ultraviolet-divergent integral.
The square-root function $S_0$ in Eq.~\eqref{bubble} is
\beq
S_0(E_\mathrm{cm} )= \sqrt{ - 2 \mu \big[ E_\mathrm{cm} -  (\varepsilon_\ast +\varepsilon) +i \epsilon \big]},
\label{S0-Ecm}
\eeq
where $E_\mathrm{cm}$ is the center-of-mass energy, 
\beq
E_\mathrm{cm} = E - P^2/(2M_X),
\label{Ecm}
\eeq
and $M_X = M_\ast + M$ is the kinetic mass of $X$.

\begin{figure}[t]
\includegraphics[width=0.95\textwidth]{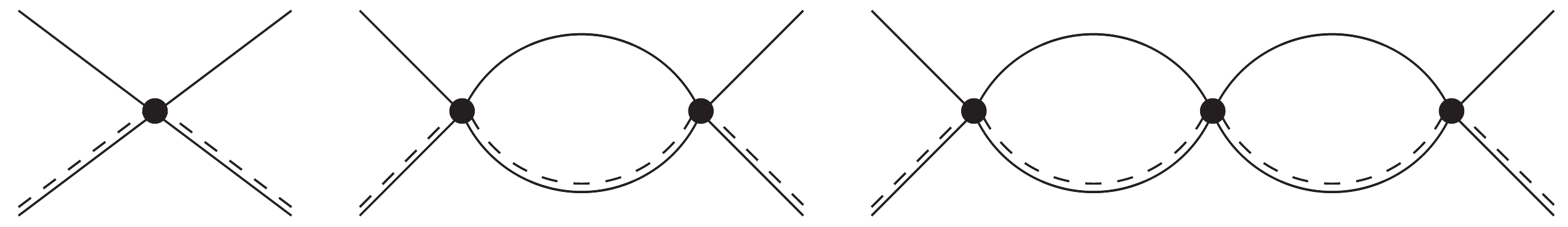}
\caption{
The first few diagrams in the geometric series of bubble-diagram corrections to the $D^\ast D$  contact vertex.
}
\label{fig:DstarDscat}
\end{figure}

The $D^\ast$ and $D$ can scatter through a contact interaction with vertex $i \,C_0$,
which is represented by the first diagram in Fig.~\ref{fig:DstarDscat}.
They can also scatter through bubble diagrams, 
 as in the second and third diagrams in Fig.~\ref{fig:DstarDscat}.
The bubble diagrams form a geometric series.
For the  molecule to be loosely bound, there must be a fine tuning of $C_0$
to nearly cancel the term proportional to $\Lambda$ in Eq.~\eqref{bubble}:
\beq
\frac{2\pi/\mu}{C_0}  =  \Lambda - \gamma_X,
\label{C-Lambda}
\eeq
where  $\gamma_X$ is small compared to $\Lambda$.
The sum of the geometric series of diagrams in Fig.~\ref{fig:DstarDscat}  is
\bqa
 i(2\pi/\mu) \bigg/ \left( \frac{2\pi/\mu}{ C_0} -  \Lambda + S_0(E_\mathrm{cm} ) \right)
= \frac{ i \, (2\pi/\mu)}{ -\gamma_X +  S_0(E_\mathrm{cm}) }.
\label{sumbubbles}
\eqa
This amplitude has a square-root branch point in $E$ at $E_\mathrm{cm} = \varepsilon_\ast + \varepsilon$.
If  $\gamma_X >0$,  it also has a nearby pole on the real axis at $E_\mathrm{cm}  = E_X$, where
\beq
E_X = \varepsilon_\ast + \varepsilon - \gamma_X^2/(2\mu) 
\label{EX-gammaX} 
\eeq
is the energy of the loosely bound molecule at zero 3-momentum.
Its binding momentum is $\gamma_X$, and
its binding energy relative to the $D^\ast D$ scattering threshold is $\gamma_X^2/(2\mu)$.

The amplitude in Eq.~\eqref{sumbubbles} is the $D^\ast D$ scattering amplitude in ZREFT at leading order.  
It can also be interpreted as the propagator for a local operator that annihilates $D$ and $D^\ast$. 
This amplitude can be obtained diagrammatically by omitting the first diagram in Fig.~\ref{fig:DstarDscat},
amputating the $D$ and $D^\ast$ legs on the remaining diagrams,
 and then taking the limit  $\Lambda \to \infty$.
The propagator differs from the scattering amplitude by the absence of the term $i\, C_0$ from
the first diagram in Fig.~\ref{fig:DstarDscat}, but that term goes to 0 in the limit $\Lambda \to \infty$.
We will refer to the amplitude on the right side of Eq.~\eqref{sumbubbles}  as the $D^\ast D$ propagator.
The corrections to the  $D^\ast D$ propagator, which is a 2-point Green function, 
are much simpler than the corrections to the $D^\ast D$ scattering amplitude, which is a  4-point Green function.

The complete $D^\ast D$ propagator includes all the corrections to the 
 sum of the bubble diagrams in Fig.~\ref{fig:DstarDscat}.
The corrections to the bubble amplitude
for the propagation of $D^\ast D$ between successive contact interactions
give an additive correction $\Sigma(E_\mathrm{cm},P)$ 
to the denominator of the  $D^\ast D$ propagator in Eq.~\eqref{sumbubbles}.
We will refer to $\Sigma$, which has dimensions of momentum, as the $D^\ast D$ self energy.
Unless there is exact Galilean invariance, $\Sigma$ can have additional dependence on 
$P$ beyond  its dependence through $E_\mathrm{cm}$.
There are $D^\ast D$-irreducible corrections to each of the contact vertices
in Fig.~\ref{fig:DstarDscat}.  The momentum-independent corrections
can be taken into account by replacing $C_0$ by a contact vertex $C_1$,  which also must be  fine tuned
to ensure that the pole remains near the threshold.

If $X$ is a loosely bound molecule, the complete $D^\ast D$  propagator must have a square-root branch point 
near  $E_\mathrm{cm} = \varepsilon_\ast + \varepsilon$, 
a pole near  $E_\mathrm{cm} = \varepsilon_\ast + \varepsilon$  when $P$ is small,
and no other nearby singularities.
The branch point comes from a function $S_1(E_\mathrm{cm},P)$ that can be chosen 
such that  the only dependence of $S_1^2$ on $E$ is an additive term $-2\mu E$.
The behavior of the complete $D^\ast D$ propagator near the branch point must then have the form
\bqa
 i\, (2\pi/\mu)\bigg/
\left( \frac{2\pi/\mu}{C_1} -  \Lambda
 +  S_0(E_\mathrm{cm}) + \Sigma(E_\mathrm{cm},P) \right)
\longrightarrow \frac{ i\, (2\pi/\mu) \, Z_X}
{ - (\gamma_X + \delta\gamma_X) +  S_1(E_\mathrm{cm},P)+\ldots},~~~
\label{Xprop-branch}
\eqa
where  $Z_X$ and $\delta\gamma_X$ are constants independent of $E_\mathrm{cm}$ and $P$.
The additional corrections to the denominator in Eq.~\eqref{Xprop-branch} are 
an expansion in powers of $S_1$ beginning at order  $S_1^2$.
The factor $Z_X$ in the numerator is determined by the condition that the 
coefficient of $S_1$ in the denominator is 1.
The constant  $\delta\gamma_X$ in the denominator is a correction to 
the binding momentum of the molecule.

\subsection{Constituent Propagator Corrections}
\label{sec:ConPropCor}

There are corrections to the propagators of $D$ and $D^\ast$ from their interactions.
The corrections to the $D$ propagator can be expressed in terms of a self energy $\Pi(E,p)$.
The effects of the self energy  include a shift $\delta \varepsilon$ in the rest energy of $D$ and
 a correction $\delta Z$ to the residue of the pole in its propagator.
If  Galilean invariance is not an exact symmetry, the  inverse kinetic mass $1/M$ 
can also be modified by a multiplicative factor $1+\zeta$.
The behavior of the complete $D$ propagator near its pole is
\beq
\frac{i}{E - \varepsilon -  \bm{p}^2/(2M) -\Pi(E,p)+ i \epsilon}
\longrightarrow
\frac{i \,(1+\delta Z)}{E - (\varepsilon+\delta \varepsilon) -(1+\zeta) \bm{p}^2/(2M) + \ldots},
\label{Dprop-pole} 
\eeq
where the additional corrections to the denominator are an expansion in powers of $p^2$ beginning at order $p^4$.
The factor $1+\delta Z$ in the numerator is determined by requiring the coefficient of $E$ in the denominator to be 1.
The corrections to the $D^\ast$ propagator can be expressed in terms of a self energy $\Pi_\ast(E,p)$.
The behavior of the complete $D^\ast$ propagator near its pole 
can be expressed in a form analogous to Eq.~\eqref{Dprop-pole} with constants
$\delta \varepsilon_\ast$, $\delta Z_\ast$, and $\zeta_\ast$.
The constants $\delta \varepsilon$, $\delta Z$, $\zeta$, $\delta \varepsilon_\ast$, $\delta Z_\ast$, 
and $\zeta_\ast$ can all be identified as corrections to parameters of ZREFT.

The form of the complete $D$ propagator in Eq.~\eqref{Dprop-pole} implies 
a change in the kinetic energy of $D$ by the factor $1+\zeta$. 
If $\zeta$ and $\zeta_\ast$ are small compared to 1, the changes in the kinetic energies
are small at small momentum $p$, but they  become increasingly large as $p$ increases.
They therefore can have a large effect on the ultraviolet behavior of 
Green functions, and this must be compensated by changes in the parameters of ZREFT.
These parameters include the $D^\ast D$  contact vertex $C_0$ in Eq.~\eqref{C-Lambda}, 
which must be  fine tuned to a new value $C_1$ to compensate 
not only for short-distance corrections to the $D$ and $D^\ast$ propagators
but also for short-distance corrections to the $D^\ast D$ contact interaction. 

In the geometric series of bubble diagrams in Fig.~\ref{fig:DstarDscat}, 
the $D$ propagator corrections can be taken into account by replacing
the $D$ propagator  by the right side of Eq.~\eqref{Dprop-pole} 
with the denominator truncated after the $p^2$ term.
The $D^*$ propagator corrections can be taken into account by making
the corresponding change in the $D^\ast$ propagator.
Changes in the contact vertex can be taken into account by replacing  $C_0$  by $C_1$.
The sum of the geometric series analogous to Eq.~\eqref{sumbubbles} is
\beq
 \frac{i(2\pi/\mu_1)}{(1+ \delta Z) (1+ \delta Z_\ast)}
   \bigg/ \left( \frac{2\pi/\mu_1}{ C_1(1+ \delta Z) (1+ \delta Z_\ast)} 
- \Lambda + \sqrt{\frac{\mu_1}{\mu}} \,S_1(E_\mathrm{cm},P ) \right),
\label{sumbubbles-1}
\eeq
where the square-root function $S_1$ is
\beq
S_1(E_\mathrm{cm} ,P)= 
\sqrt{ - 2 \mu \big[ E_\mathrm{cm} -  (\varepsilon_\ast + \varepsilon) 
- (\delta\varepsilon_\ast +\delta\varepsilon)  - \zeta_X P^2/(2M_X) + i \epsilon\big]}.
\label{S1-E}
\eeq
The modified reduced mass $\mu_1$ is obtained from $\mu$ by making the substitutions
 $M \to M/(1+\zeta)$ and $M_\ast \to M_\ast/(1+\zeta_\ast)$:
\beq
\frac{\mu}{\mu_1} = 1 + \zeta  \frac{M_\ast}{M_X} + \zeta_\ast \frac{M}{M_X}.
\label{mu1} 
\eeq
The constant  $\zeta_X$ is
\beq
\zeta_X =
 \frac{\zeta_\ast (1 + \zeta)M_\ast + \zeta (1 + \zeta_\ast)M}{(1 + \zeta)M_\ast + (1 + \zeta_\ast)M}.
\label{zetaX} 
\eeq
The $D^\ast D$ contact vertex $C_1$ required to compensate for constituent propagator corrections
can be determined by identifying  the ultraviolet-divergent terms $-\Lambda$ in the denominators of 
 Eqs.~\eqref{sumbubbles} and \eqref{sumbubbles-1}.  The first term in the denominator of 
Eq.~\eqref{sumbubbles-1} must then be identical to the corresponding constant $(2 \pi/\mu)/C_0$ in Eq.~\eqref{sumbubbles}.
The required contact vertex can be expressed as
\beq
\frac{2\pi/\mu}{C_1} = \frac{2\pi/\mu}{C_0+\delta C}  \frac{(1+ \delta Z)(1+ \delta Z_\ast)}{ \mu/\mu_1}.
\label{C1-C0} 
\eeq
We have replaced $C_0$ by $C_0+\delta C$ to allow for additional  corrections to 
the contact vertex required to compensate for the effects of interactions between constituents.

The $D^\ast D$ self  energy $\Sigma(E_\mathrm{cm},P)$ in Eq.~\eqref{Xprop-branch}
 must include terms that change $S_0(E_\mathrm{cm} )$ in Eq.~\eqref{S0-Ecm}
into $S_1(E_\mathrm{cm},P)$ in Eq.~\eqref{S1-E}.
It must therefore include terms that are singular at the branch point of $S_0(E_\mathrm{cm} )$.
The expansion of  $S_1(E_\mathrm{cm},P)$ to first order in
$\delta\varepsilon_\ast$, $\delta\varepsilon$, and $\zeta_X$ is
\beq
S_1(E_\mathrm{cm} ,P) \approx S_0(E_\mathrm{cm}) 
+\mu (\delta\varepsilon_\ast + \delta\varepsilon) \frac{1}{S_0(E_\mathrm{cm})} 
+ \zeta_X \frac{ \mu}{2M_X} \frac{P^2}{S_0(E_\mathrm{cm})}  .
\label{S1-S0}
\eeq
At successively higher orders, there are increasingly singular terms in the form of
increasing powers of $1/S_0(E_\mathrm{cm} )$.
These singular terms must be resummed to all orders to change the function $S_0(E_\mathrm{cm})$
into  $S_1(E_\mathrm{cm} ,P)$.


\section{Charm mesons in a Pion Gas}
\label{sec:Charm/Pi}

In this Section, we introduce notation for some of the properties of charm mesons and pions.
We describe the pion gas that can be produced by a heavy-ion collision,
and we specify the momentum distribution for pions in the pion gas.
We also define thermal averages that appear in the
self energies of the charm mesons and of the charm-meson pair in the pion gas.

\subsection{Charm Mesons and Pions}

We consider a loosely bound charm-meson molecule that is a bound state of
a vector charm meson and a pseudoscalar charm meson.
We denote the molecule by $X$ and the charm mesons by $D^{\ast a}$  and $D^b$,
where $a$ and $b$ are charm-meson flavor indices.
(These charm mesons may contain a charm antiquark instead of a charm quark.)
In the case where $X$ is $T_{cc}^+(3785)$, the constituents are $D^{*+}D^0$, 
which are the $(a,b) = (1,2)$ members of a charm-meson isospin doublet.
In the case where $X$ is $X(3872)$, the constituents  are $(D^{*0} \bar D^0 + \bar D^{*0} D^0 )/\sqrt2$,
which is a superposition of the $(a,b) = (2,1)$ and $(1,2)$ members of charm-meson isospin doublets.

We treat pions as relativistic particles.
We denote the mass of the pion with flavor $i$, which can be +, 0, or $-$, by $m_{\pi i}$.
The  flavor-averaged pion mass is $m_\pi = 138.0$\,MeV. 
We represent the propagator for $\pi$ by a dashed line.
The self interactions of pions in chiral effective field theory ($\chi$EFT) at leading order
are determined by the pion decay constant $f_\pi = 131.7$\,MeV.

We treat  charm mesons as nonrelativistic particles.
We denote the masses of $D^{\ast c}$ and   $D^d$ by $M_{\ast c}$ and  $M_d$. 
We decompose  them into kinetic masses $M_\ast$ and $M$
and rest energies $\varepsilon_{\ast c}$ and $\varepsilon_d$: 
$M_{\ast c} = M_\ast+\varepsilon_{\ast c}$ and  $M_d = M +\varepsilon_d$.
Convenient choices for $M$ and $M_\ast$ are the isospin averages of the charm-meson masses:
$M= (M_+ + M_0)/2$, $M_\ast = (M_{\ast +} + M_{\ast 0})/2$.
The rest energies  $\varepsilon_{\ast c}$ and $\varepsilon_d$ are then comparable to isospin splittings.
The charm-meson mass differences $\Delta_{cd} = M_{\ast c} - M_d$  differ from  $m_\pi$
by amounts comparable to isospin splittings.
The value of $\Delta_{cd}$ averaged over the four $D^\ast \to D$ transitions is $\Delta = M_\ast - M = 141.3$\,MeV.
We represent the propagator for $D$ by a solid line.
Since the mass of $D^\ast$ is close to the sum of the masses of $D\pi$,
we represent the propagator for $D^\ast$  by a double line (solid+dashed).
We take the vertices for interactions between charm mesons and pions 
to be the nonrelativistic form of those in heavy-hadron $\chi$EFT at leading order.
The interaction parameters are $f_\pi$ and a dimensionless coupling constant $g_\pi = 0.520$, 
whose value is determined from the decay rate for $D^{\ast +} \to D^0 \pi^+$.
The Feynman rules used in our calculations are given in Appendix~\ref{app:Feynman}.

We treat the loosely bound charm-meson molecule using a ZREFT for the nonrelativistic charm mesons
analogous to that described in Sec.~\ref{subsec:PairProp}.
The energy of $X$ relative to the $D^{\ast a}D^b$ threshold
can be expressed as $\varepsilon_X = -\gamma_X^2/(2\mu)$,
where $\gamma_X$ is the binding momentum and $\mu = M M_\ast/(M_\ast+M)$
is the reduced kinetic mass.  The assumption that $X$ is loosely bound
 is equivalent to $|\varepsilon_X| \ll m_\pi^2/M$ or  $|\gamma_X| \ll m_\pi$.

The most accurate determinations of the mass and width of the $X(3872)$ resonance 
have been made by the LHCb collaboration  \cite{LHCb:2020xds,LHCb:2020fvo}.
A fit using  a Breit-Wigner line shape gives $E_\mathrm{BW}=-0.07 \pm 0.12$\,MeV  
for the energy relative to the $D^{*0} \bar D^0$ threshold 
and  $\Gamma_\mathrm{BW}=1.19\pm0.19$\,MeV  for the width.
Using a Flatt\'e parameterization  that takes into account the nearby $D^{*0} \bar D^0$ threshold
and the $D^{*0}$ width,
 the best fit to the $X$ line shape in the $J/\psi\, \pi^+\pi^-$ decay channel gives $(+25 - 140\, i)$\,keV 
for the pole energy  relative to the  threshold \cite{LHCb:2020xds}. 
The complex binding momentum $\gamma_X$  can be determined by
identifying this energy with $-\gamma_X^2/(2\mu) -i\,\Gamma_{*0}/2$,
where $\mu$ is  the $D^{*0} \bar D^0$ reduced mass
and $\Gamma_{\ast 0} =55.4$\,keV is the predicted  $D^{*0}$ decay width.
The solution with a positive real part, corresponding to a bound state,
 is $\gamma_X = (9.3+11.6\, i)$\,MeV. 
The positive imaginary part takes into account short-distance decay modes of $X$,
such as $J/\psi\, \pi^+\pi^-$ and $J/\psi\,  \pi^+\pi^-\pi^0$.
These decays account for most of the width $\Gamma_X = 280$\,keV 
determined by the pole energy of $X$ in Ref.~\cite{LHCb:2020xds}.

Accurate determinations of the mass and width of the  $T_{cc}^+(3875)$  resonance 
have been made by the LHCb collaboration  \cite{LHCb:2021auc}.
A fit to the  $T_{cc}^+$ line shape  in the $D^0 D^0 \pi^+$ decay channel using  a Breit-Wigner line shape gives
$E_\mathrm{BW} = -273 \pm 61$\,keV for the energy relative to the $D^{*+} D^0$ threshold 
and $\Gamma_\mathrm{BW} = 410  \pm 165$\,keV for the width. 
Using a unitarized model that takes into account the nearby $D^{*+} D^0$ threshold, 
the central value of the pole energy  relative to the threshold is $(-360 - 24\, i)$\,keV \cite{LHCb:2021auc}. 
The complex binding momentum $\gamma_T$ can be determined by
identifying the pole energy with $-\gamma_T^2/(2\mu) -i\,\Gamma_{*+}/2$,
where $\mu$ is  the $D^{*+} D^0$ reduced mass
and $\Gamma_{\ast +} = 83.4$\,keV is the measured  $D^{*+}$ decay width.
The solution with a positive real part, corresponding to a bound state, is  $\gamma_T = (26.4 - 0.6\,i)$\,MeV.
The small negative imaginary part takes into account decays from the small $D^{*0} D^+$
component of the bound state.
The $D^{*0} D^+$ scattering threshold is higher than  the  $D^{*+} D^0$ threshold by only 1.4\,MeV.
The  $D^{*0} D^+$ component can decay into $D^+ D^0 \pi^0$ through a decay of the $D^{\ast 0}$ constituent,
and this interferes destructively with the decay $D^{\ast +} \to D^+\pi^0$ from the dominant $D^{*+} D^0$ component.
Estimates of the decay width of $T_{cc}^+$ that take into account the coupled 
$D^{*+} D^0$ and $D^{*0} D^+$ channels range from 36 to 78\,keV
\cite{Meng:2021jnw,Ling:2021bir,Feijoo:2021ppq,Yan:2021wdl,Dai:2021wxi,Fleming:2021wmk,Albaladejo:2021vln,Du:2021zzh},
all of which are smaller than the width $\Gamma_{\ast +}$ of $D^{\ast +}$.
The  width $\Gamma_T = 48$\,keV determined by the pole energy of $T_{cc}^+$ in Ref.~\cite{LHCb:2021auc}
is a little more than half of $\Gamma_{\ast +}$.

\subsection{Pion Gas}
\label{sec:PionGas}

A central relativistic heavy-ion collision can produce a region of quark-gluon plasma 
that expands and cools while in thermal equilibrium.
When it reaches the hadronization temperature of 156\,MeV,
it undergoes a transition to a hadron resonance gas, 
which continues to expand and cool while in thermal equilibrium until kinetic freeze-out. 
For Pb-Pb collisions at the LHC, the kinetic freeze-out temperature $T_\mathrm{kf}$ is 
estimated to be 115\,MeV \cite{ExHIC:2017smd}. 
At temperatures $T$ near $T_\mathrm{kf}$, the hadron resonance gas can be approximated by a pion gas.
The momentum distribution of the pions is a Bose-Einstein distribution
$1/(e^{\beta \omega_q} -1)$, where $\omega_q = \sqrt{m_\pi^2 + q^2}$ and $\beta = 1/T$.
If the pions are in thermal equilibrium at temperature $T$, 
the number density for each of the three pions is
\beq
\mathfrak{n}_\pi^\mathrm{(eq)} =  \int \!\!\frac{d^3q}{(2\pi)^3} \frac{1}{e^{\beta \omega_q} -1}.
\label{npieq}
\eeq
At $T_\mathrm{kf}=115$\,MeV,
the equilibrium pion number density is  $1/(3.9~\mathrm{fm})^3$. 

After  kinetic freeze-out, the hadron gas is no longer in thermal equilibrium, 
but it continues to expand by the free-streaming of hadrons.
It can be approximated by a pion gas with a decreasing number density $\mathfrak{n}_\pi$ 
and a fixed temperature $T_\mathrm{kf}$.
(Such a system could also be produced by the isothermal expansion of a pion gas
that was in thermal equilibrium at temperature  $T_\mathrm{kf}$.)
The pion momentum distribution in both the expanding and cooling pion gas  before kinetic freeze-out
and  the expanding pion gas after kinetic freeze-out can be described by
\beq
\mathfrak{f}_\pi(\omega_q) = \frac{\mathfrak{n}_\pi }{\mathfrak{n}_\pi^\mathrm{(eq)} }
\frac{1}{e^{\beta \omega_q} -1}.
\label{fpi-n}
\eeq
Before kinetic freeze-out, $T$ decreases with time 
and $\mathfrak{n}_\pi =\mathfrak{n}_\pi^\mathrm{(eq)}$ is determined by $T$ using Eq.~\eqref{npieq}.
After kinetic freeze-out, $\mathfrak{n}_\pi$ decreases with time 
and the temperature remains fixed at $T =T_\mathrm{kf}$.
A dimensionless number that characterizes the size of the effects of thermal pions is
$\mathfrak{n}_\pi/(f_\pi^2 m_\pi)$. At $T_\mathrm{kf}=115$\,MeV, this number is small:
 $\mathfrak{n}_\pi^\mathrm{(eq)}/(f_\pi^2 m_\pi) = 0.052$. 
 
There have been many previous discussions of the production of exotic heavy hadrons in heavy ion collisions
\cite{ExHIC:2017smd,ExHIC:2010gcb,ExHIC:2011say,Hong:2018mpk,Fontoura:2019opw,Hu:2021gdg,Hu:2023hrn},
including some that focused on $X(3872)$ and $T_{cc}^+(3875)$
\cite{Cho:2013rpa,MartinezTorres:2014son,Zhang:2020dwn,Wu:2020zbx,Chen:2021akx,Abreu:2022lfy,Yun:2022evm,Guo:2023dwf}.
Most treatments of hadronic molecules have taken into account their size,
but are otherwise uninformed about the physics of loosely bound molecules. 
There have been previous calculations of the thermal mass shift and the thermal width of $X(3872)$ 
in a hadron gas \cite{Cleven:2019cre,Montana:2022inz}.
The primary goal of this paper is  the calculation of  these properties for
a loosely bound charm-meson molecule in a pion gas.

We proceed to summarize the important energy and momentum scales of the system.
There is a large energy scale set by the temperature $T$, which is comparable to $m_\pi$.
The typical momentum $q$ and energy $\omega_q$ of a pion are order $m_\pi$.
There is a much larger energy scale set by the charm-meson masses $M_\ast$ and $M$.
Isospin splittings and the $D^\ast$-to-$D\pi$ mass differences $\Delta_{cd} - m_{\pi i}$
are all smaller than $m_\pi^2/M \sim10$\,MeV. The strong  inequality $m_\pi^2/M \ll m_\pi$
allows a charm-meson propagator with momentum of order $m_\pi$ 
to be expanded in powers of its kinetic energy, which is order $m_\pi^2/M$, and 
in powers of isospin splittings, which we take to also be order $m_\pi^2/M$.
We refer to this expansion as the {\it heavy-meson expansion}.
In an amplitude for the propagation of a charm-meson pair, the heavy-meson expansion 
produces integrals over the relative momentum of a charm-meson pair that are ultraviolet divergent.
These integrals can be defined by imposing the ultraviolet momentum cutoff $\Lambda$
of ZREFT, which we take to be order $m_\pi$.
The binding momentum $\gamma_X$ of the charm-meson molecule is assumed to be much smaller than $m_\pi$, 
which implies that the binding energy $\gamma_X^2/(2 \mu)$ is much much smaller than $m_\pi^2/M $.
The hierarchy of energy scales can be succinctly summarized as
\begin{equation*}
     \gamma_X^2/M \ll |\Delta - m_\pi|,  m_\pi^2/M \ll m_\pi, T \ll M,M_\ast .
\end{equation*}

\subsection{Thermal Averages}
\label{ThermAve}

We use angular brackets to denote the average over the Bose-Einstein momentum distribution of a pion.
The thermal average of a function $F(\bm{q})$ of the pion momentum is
\beq
 \big\langle F(\bm{q}) \big\rangle = 
 \int \frac{d^3q}{(2\pi)^3}\,  \mathfrak{f}_\pi(\omega_q) \, F(\bm{q}) 
\bigg/  \int \frac{d^3q}{(2\pi)^3}\,  \mathfrak{f}_\pi(\omega_q).
\label{<F>}
\eeq
The thermal average depends on the temperature $T$ but not on the pion number density $\mathfrak{n}_\pi$.
It may be sensitive to the flavor $i$ of the pion,
in which case the pion energy in Eq.~\eqref {<F>}
should be replaced by $\omega_{iq}  = \sqrt{m_{\pi i}^2+q^2}$.

The propagators of charm mesons in the pion gas involve thermal averages 
over the pion momentum distribution.
Some of the  thermal averages have the simple form 
$\langle (q^2)^n/\omega_q^m \rangle$,
but other thermal averages are more sensitive to isospin splittings.
The thermal average that appears in the charm-meson self energies at leading order (LO) 
in the heavy-meson expansion is
\beq
\mathcal{F}_{cd} = 
\left \langle \frac{q^2}{\omega_{cdq}\, (\omega_{cdq}^2 - \Delta_{cd}^2 + i \epsilon)} \right\rangle,
\label{Fcd}
\eeq
where $\Delta_{cd}$ is the $D^{\ast c}$-$D^d$ splitting, $\omega_{cdq} = \sqrt{m_{\pi cd}^2 + q^2}$,
and $m_{\pi cd}$ is an alternative notation 
for the mass $m_{\pi i}$ of the pion produced by the transition $D^{\ast c} \to D^d \pi^i$.
The thermal averages that appear in the charm-meson self energies 
at next-to-leading order (NLO) in the heavy-meson expansion  include
\beq
\mathcal{G}_{n,cd} = 
 \left \langle
\frac{(q^2)^n \, (\omega_{cdq}^2 + \Delta_{cd}^2)}{\omega_{cdq}\, (\omega_{cdq}^2 - \Delta_{cd}^2 + i \epsilon)^2} \right\rangle,
\label{Gcd,n}
\eeq
where $n$ is 1 or 2, and
\beq
\mathcal{H}_{cd} = 
 \left\langle  \frac{q^2 (\omega_{cdq}^2+ \Delta_{cd}^2)}
{\omega_{cdq}\, [ (\omega_{cdq}^2 - \Delta_{cd}^2)^2 + \epsilon^2]} \right\rangle .
\label{Hcd}
\eeq

The thermal averages in Eqs.~\eqref{Fcd} and \eqref{Gcd,n}
can be defined by the limit $\epsilon \to 0^+$.
In  Appendix~\ref{app:IntMom},
their real parts are expressed as principle-value integrals over a momentum variable
and their imaginary parts are evaluated analytically.
The thermal averages can be expanded in powers of isospin splittings divided by $m_\pi$.
The leading terms in the expansions of their real and imaginary parts are given in 
 Appendix~\ref{sec:ExpandSplitting}.
The thermal average $\mathcal{H}_{cd}$ in Eq.~\eqref{Hcd} 
 has terms that diverge as $1/\epsilon$ as $\epsilon \to 0$.
The divergence can be regularized by taking into account the widths of the charm mesons
 as described in Appendix~\ref{sec:Hcd}.


\section{Charm-meson Self Energies}
\label{sec:CharmMesons}

In this Section, we calculate the self energies of pseudoscalar and vector charm mesons in the pion gas
to NLO in the heavy-meson expansion.
The self energies are used to identify thermal corrections to parameters of ZREFT.

\subsection{Corrections from Pion Forward Scattering}
\label{ThermCor}

In a system of pions and nonrelativistic charm mesons with temperature $T$,
the charm meson has a typical kinetic energy of order $T$ and a typical momentum of order $\sqrt{MT}$.
If $T$ is  order $m_\pi$, the typical charm-meson kinetic energy is order $m_\pi$.
However, for simplicity we will calculate the self energies of charm mesons 
with a parametrically smaller kinetic energy $E$ of order $m_\pi^2/M$.
The  self energies are calculated at leading order in the pion interactions 
and to NLO  in the heavy-meson expansion.
\begin{figure}[t]
\includegraphics[width=0.65\textwidth]{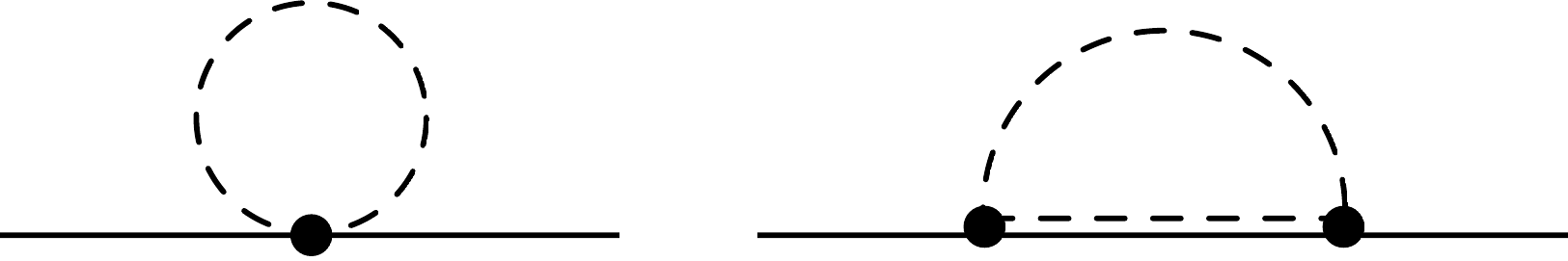}
\caption{
One-loop diagrams for the D self energy in thermal field theory. 
}
\label{fig:Dselfenergy-new}
\end{figure}

The effect of the pion gas on the propagation of charm mesons
can be taken into account using thermal field theory.
In the imaginary-time formalism, the self energy of a $D$ meson at leading order
is the sum of the two one-loop diagrams in Fig.~\ref{fig:Dselfenergy-new}.
The second diagram, which has an internal pion line and an internal charm-meson line,
includes a zero-temperature part, a thermal part from the coherent forward scattering of an on-shell pion,
and a thermal part from the coherent forward scattering of an on-shell charm meson.
We ignore the zero-temperature part, because it is taken into account in the parameters of the ZREFT.
We ignore the thermal charm-meson part, because the thermal distribution of the charm meson
has an exponential suppression factor $\exp(-M/T)$.
The thermal pion part can alternatively be represented by the forward-scattering 
diagram in Fig.~\ref{fig:Dselfenergy}.
An on-shell pion emerges from one open circle with momentum $\bm{q}$ and flavor $i$ and 
is scattered back into the other open circle with the same momentum and  flavor.
The amplitudes must be added coherently by multiplying them by 
$\mathfrak{f}_\pi(\omega_q)/(2\omega_q)$,
where $\mathfrak{f}_\pi(\omega_q)$ is the pion momentum distribution, 
integrating over $\bm{q}$  with measure $d^3q/(2\pi)^3$, and summing over the  three flavors $i$.

\subsection{Pseudoscalar charm meson}
\label{PiD}

The complete propagator for a pseudoscalar charm meson $D^b$ 
with energy $E$ (relative to its kinetic mass $M$) and momentum  $\bm{p}$
has the form on the left side of Eq.~\eqref{Dprop-pole},  with rest energy $ \varepsilon_b$ 
and self energy  $\Pi_b(E,p)$.
The expansion of the  denominator near the pole in $E$ has   the form
\beq
E - \varepsilon_b - \frac{p^2}{2M} - \Pi_b(E,p) = 
(1+\delta Z_b)^{-1} \left[ E - (\varepsilon_b + \delta \varepsilon_b) - (1 +\zeta_b) \frac{p^2}{2M} + \ldots \right],
\label{Dbpropexpand}
\eeq
where $\delta \varepsilon_b$ is the thermal rest energy, 
$\delta Z_b$ is the change in the residue of the pole, 
and $\zeta_b$ determines the change in the kinetic mass.
The constants $\delta \varepsilon_b$, $\delta Z_b$, and $\zeta_b$
can be identified as corrections to parameters of ZREFT.

\begin{figure}[t]
\includegraphics[width=0.6\textwidth]{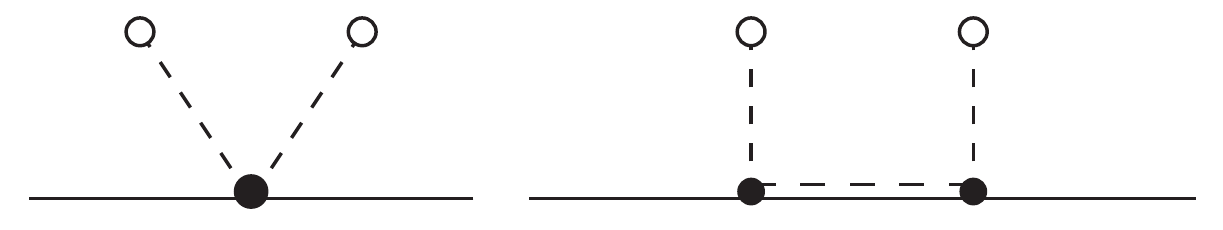}
\caption{
Diagrams for the $D$ self energy  from pion forward scattering.
The second diagram is summed over the two directions for the routing of the pion momentum.
The contribution to $-i \, \Pi(E,p)$
is the sum of the diagrams with $D$ legs and pion legs amputated, 
weighted by $\mathfrak{f}_\pi(\omega_q)/(2\omega_q)$,
integrated over $\bm{q}$, and summed over the pion flavor $i$.
}
\label{fig:Dselfenergy}
\end{figure}

At leading order in the  pion interactions, the thermal contribution to the  $D$ self energy
comes from pion forward scattering through the diagrams  in Fig.~\ref{fig:Dselfenergy}.
The contributions from the first diagram cancel upon summing over the pion flavor.
In the contributions from the second diagram,
the sum over the pion  flavor $i$ of the Clebsch-Gordan factors from the pion vertices
can be evaluated using a simple identity
for Pauli matrices: $\sum_i  \big|\sigma^i_{cd}\big|^2 = 2 - \delta_{cd}$,
where $c$ and $d$ are flavor indices for charm mesons.
The self energy from the second diagram in Fig.~\ref{fig:Dselfenergy} is
\bqa
\Pi_b(E,p) &=& \frac{g_\pi^2}{4 f_\pi^2} \, \mathfrak{n}_\pi  \sum_c (2 - \delta _{cb} ) 
\left \langle \frac{q^2}{\omega_{cbq}} \sum_\pm 
\frac{1}{E-  \varepsilon_b \pm \omega_{cbq} - \Delta_{cb} - (\bm{p} \!\pm\! \bm{q})^2/(2M_\ast) + i \epsilon}
\right\rangle,~~~
\label{PiDb}
\eqa
where the sum is over the flavor of the intermediate $D^{\ast c}$,
$\Delta_{cb}$ is the $D^{\ast c}$-$D^b$ mass difference, 
$\omega_{cbq} = \sqrt{m_{\pi cb}^2 + q^2}$,
and $m_{\pi cb}$ is the mass of the pion produced by the transition $D^{\ast c} \to D^b \pi$.
The angular brackets, which are defined in Eq.~\eqref{<F>},
indicate the average over the Bose-Einstein  distribution for the momentum $\bm{q}$ of the pion with flavor $cb$. 

The terms  $\omega_{cbq}$ and $\Delta_{cb}$ 
in the denominator of the self energy in Eq.~\eqref{PiDb} are order $m_\pi$.
The heavy-meson expansion is obtained by expanding Eq.~\eqref{PiDb}
 in powers of the small energies $E - \varepsilon_b$ and $(\bm{p} \pm \bm{q})^2/(2M_\ast)$,
 which we consider to be order $m_\pi^2/M$.
Terms with odd powers of $\bm{q}$ vanish after averaging over the directions of the pion momentum.
The  LO term is a constant independent of $E$ and $p$:
\beq
\Pi_b^\mathrm{(LO)} = 
\frac{g_\pi^2}{2 f_\pi^2}  \, \mathfrak{n}_\pi \sum_c (2 - \delta _{cb} )\,  \Delta_{cb} \, \mathcal{F}_{cb},
\label{PiDb-LO}
\eeq
where $\mathcal{F}_{cb}$ is the thermal average defined in Eq.~\eqref{Fcd}.
The  NLO term is a linear  function of $E$ and $p^2$:
\bqa
\Pi_b^\mathrm{(NLO)}(E,p) &=&- \frac{g_\pi^2}{2 f_\pi^2} \, \mathfrak{n}_\pi  \sum_c (2 - \delta _{cb} ) 
\left[ \left(E - \varepsilon_b - \frac{p^2}{2M_\ast}\right)  \mathcal{G}_{1,cb} 
 - \frac{1}{2M_\ast}   \mathcal{G}_{2,cb} \right],
\label{PiDb-NLO}
\eqa
where $ \mathcal{G}_{n,cb}$ is the thermal average defined in Eq.~\eqref{Gcd,n}.
We can read off the corrections $\delta  \varepsilon_b$, $\delta Z_b$, and  $\zeta_b$
 defined by the expansion in Eq.~\eqref{Dbpropexpand}:
\begin{subequations}
\bqa
\delta  \varepsilon_b &=& \frac{g_\pi^2}{2f_\pi^2}  \, \mathfrak{n}_\pi \sum_c (2 - \delta _{cb} )
\left(   \Delta_{cb}\, \mathcal{F}_{cb} + \frac{1}{2M_\ast}  \mathcal{G}_{2,cb} \right),
\label{deltaepsilonDb}
\\
\delta Z_b &=& - \frac{g_\pi^2}{2 f_\pi^2}  \, \mathfrak{n}_\pi \sum_c (2 - \delta _{cb} ) \,
\mathcal{G}_{1,cb},
\label{deltaZDb}
\\
\zeta_b &=& 
- \frac{g_\pi^2 \, \Delta}{2 f_\pi^2 M_\ast}  \, \mathfrak{n}_\pi \sum_c (2 - \delta _{cb} ) \,\mathcal{G}_{1,cb},
\label{deltazetaDb}
\eqa
\label{deltaDb}%
\end{subequations}
where $\Delta = M_\ast - M$.
For a pion gas at $T_\mathrm{kf} = 115$\,MeV, the thermal rest energies for $D^+$ and $D^0$  
are $\delta  \varepsilon_+ = (1.60 - 0.018 \,i)$\,MeV 
and $\delta  \varepsilon_0 = (1.77 - 0.065 \,i)$\,MeV. 
Their real parts are the thermal mass shifts.
The thermal widths are their imaginary parts multiplied by $-2$: $\delta \Gamma_b = -2\, \mathrm{Im}[\varepsilon_b]$.
The changes in the residue factors are
 $\delta Z_+  = -0.022 + 0.004\, i$ and $\delta Z_0  =-0.024 + 0.013\, i$. 
The correction factor $\zeta_b$ differs from $\delta Z_b$ by the small factor  $\Delta/M_\ast$.

The $D$ self-energy in Eq.~\eqref{PiDb}
comes from P-wave pion interactions  through the second diagram of Fig.~\ref{fig:Dselfenergy}.
The contribution from S-wave pion interactions through  the first diagram in Fig.~\ref{fig:Dselfenergy} is 0.
This zero implies a cancellation between the contributions from $\pi D$ scattering 
in the channels with total isospin $\tfrac12$ and $\tfrac32$.
The S-wave contribution to the $D$ self energy at leading order in the chiral expansion can be expressed as 
\beq
\Pi_S^{(\mathrm{LO})} = \frac{1}{f_\pi^2}\,  \mathfrak{n}_\pi \,  \big( 1 - 1 \big),
\label{PiS}
\eeq
where the two canceling terms in the last factor come from total isospin $\tfrac12$ and $\tfrac32$.
At higher orders in the chiral expansion, the canceling factor $(1-1)$ is replaced by 
terms suppressed by powers of $T/ (4 \pi f_\pi$) or $m_\pi/(4 \pi f_\pi)$.
For a pion gas at $T_\mathrm{kf} = 115$~MeV, the canceling S-wave contributions 
to the $D$ thermal mass shift at leading order in Eq.~\eqref{PiS} are $\pm 7.2$~MeV. 
These canceling S-wave contributions are larger than the P-wave contribution in Eq.~\eqref{deltaepsilonDb}.
This suggests that S-wave contributions
suppressed by powers of $m_\pi/(4 \pi f_\pi)$ or $T/(4 \pi f_\pi)$ could be significant.

The exact contribution to the $D$ self-energy from the coherent forward scattering 
of low-energy pions in the S-wave channel can be expressed as
\beq
\Pi_S = \frac{2\pi}{\mu} \mathfrak{n}_\pi  \left( a_{\pi D}^{(1/2)} + 2 \, a_{\pi D}^{(3/2)} \right),
\eeq
where $\mu= M\,m_\pi/(M+m_\pi)$ is the $\pi D$ reduced mass 
and $a_{\pi D}^{(1/2)}$ and $a_{\pi D}^{(3/2)}$ are the S-wave scattering lengths 
in the channels with  total isospin $\tfrac12$ and $\tfrac32$.
At leading order in the chiral expansion, the scattering lengths are
$a_{\pi D}^{(1/2)} =\mu/(2 \pi f_\pi^2)$ and $a_{\pi D}^{(3/2)} = -\mu/(4 \pi f_\pi^2)$. 
These leading-order scattering lengths are $a_{\pi D}^{(1/2)} = 0.233$\,fm and $a_{\pi D}^{(3/2)} = -0.116$\,fm.
The  scattering length $a_{\pi D}^{(3/2)}$ has been calculated using lattice QCD \cite{Liu:2008rza,Liu:2012zya}.
The extrapolation to the physical pion mass gives $a_{\pi D}^{(3/2)} = -0.16(4)$~fm \cite{Liu:2008rza}.
The  scattering length $a_{\pi D}^{(1/2)}$ has been calculated using lattice QCD \cite{Mohler:2012na,Moir:2016srx},
but it has not been extrapolated to the physical pion mass.

There have been several calculations of the S-wave $\pi D$ scattering lengths
beyond leading order in the chiral expansion \cite{Guo:2009ct,Liu:2009uz,Geng:2010vw,Liu:2012zya}.
The S-wave scattering lengths for all the pseudoscalar light mesons
and pseudoscalar charm mesons
through NLO  in the chiral expansion are given in Ref.~\cite{Guo:2009ct}.
The canceling factor $(1-1)$ in the S-wave self energy at LO in Eq.~\eqref{PiS}
is replaced at NLO by $(h_1-3h_3 - 6 h_5^\prime) m_\pi/M$,
where $h_1$, $h_3$, and $h_5^\prime$ are dimensionless low-energy constants.
The constant $h_1$ is determined by charm-meson mass splittings,
but $h_3$ and $h_5^\prime$ have not been determined definitively.
The calculations beyond leading order have used various prescriptions
to fix the unknown low-energy constants.
The resulting factors replacing $(1-1)$ in Eq.~\eqref{PiS} range from $-0.36$ to +0.67  \cite{Guo:2009ct,Liu:2009uz,Geng:2010vw,Liu:2012zya}.
The largest values were obtained using unitarized chiral perturbation theory
constrained to reproduce the mass of the strange charm meson $D^\ast_{s0}(2317)$ \cite{Guo:2009ct,Liu:2012zya}.
All the results for the S-wave contribution to the $D$ thermal mass shift are smaller than
the $I=\tfrac12$ term at LO, which is $ \mathfrak{n}_\pi/f_\pi^2$.

\subsection{Vector charm meson}
\label{PiD*}

The complete propagator for a vector charm meson $D^{\ast a}$ 
with energy $E$ (relative to its kinetic mass $M_\ast$) and momentum  $\bm{p}$ 
has the form on the left side of Eq.~\eqref{Dprop-pole},  with rest energy $ \varepsilon_{\ast a}$,
kinetic mass $M_\ast$, and self energy  $\Pi_{\ast b}(E,p)$.
The behavior of the  denominator near the pole in $E$ has  the form
\beq
E - \varepsilon_{\ast a} - \frac{p^2}{2M_\ast} - \Pi_{\ast a}(E,p) = 
(1+\delta Z_{\ast a})^{-1} \left[ E - (\varepsilon_{\ast a} + \delta \varepsilon_{\ast a}) 
- (1 +\zeta_{\ast a}) \frac{p^2}{2M_\ast} + \ldots \right],
\label{D*apropexpand}
\eeq
where $\delta \varepsilon_{\ast a}$ is the  thermal rest energy, 
 $\delta Z_{\ast a}$ is the change in the residue of the pole, 
and $\zeta_{\ast a}$ determines the  change in the kinetic mass.
The constants $\delta \varepsilon_{\ast a}$, $\delta Z_{\ast a}$, and $\zeta_{\ast a}$
can be identified as corrections to parameters of ZREFT.

\begin{figure}[t]
\includegraphics[width=0.96\textwidth]{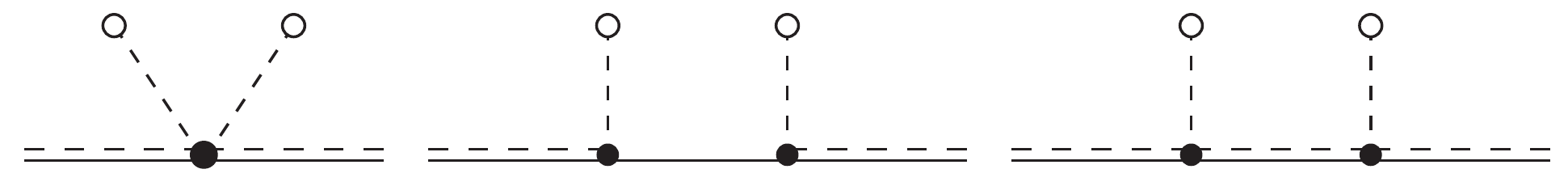}
\caption{
Diagrams for the $D^\ast$ self energy from pion forward scattering.
The last two diagrams are summed over the two directions for routing the pion momentum.
The contribution to $-i \,\Pi_\ast^{mn}(E,p)$
is the sum of the diagrams with $D^\ast$ legs and pion legs amputated, 
weighted by $\mathfrak{f}_\pi(\omega_q)/(2\omega_q)$,
integrated over $\bm{q}$, and summed over the pion flavor $i$.
}
\label{fig:D*selfenergy}
\end{figure}

At leading order in the  pion interactions, the thermal contribution to the $D^\ast$  self energy
comes from pion forward scattering through the diagrams in Fig.~\ref{fig:D*selfenergy}.
The contributions from the first diagram cancel upon summing over the pion flavor.
The self energy tensor from the last two diagrams is
\bqa
\Pi_{\ast a}^{mn}(E,p) &=& \frac{g_\pi^2}{4 f_\pi^2}  \, \mathfrak{n}_\pi \left\{ \sum_d (2 \!-\! \delta _{ad} ) 
 \left \langle \frac{q^m q^n}{\omega_{adq}} \sum_\pm
\frac{1}{E - \varepsilon_{\ast a} \pm \omega_{adq} + \Delta_{ad} - (\bm{p} \!\pm\! \bm{q})^2/(2M) + i \epsilon}
\right\rangle \right.
\nonumber\\
&&\hspace{-0.5cm}  \left.
+ \sum_c (2 \!-\! \delta _{ac} ) 
\left \langle \frac{q^2 \delta^{mn} \!-\! q^m q^n}{\omega_{acq}} \sum_\pm
\frac{1}{E - \varepsilon_{\ast c} \pm \omega_{acq}- (\bm{p} \!\pm\!\bm{q})^2/(2M_\ast) + i \epsilon}
\right\rangle \right\},
\label{PiD*aij}
\eqa
where the sums are over the flavors of the intermediate $D^d$ and the intermediate $D^{\ast c}$
and $\Delta_{ad}$ is the $D^{\ast a}$-$D^d$ mass difference.
The angular brackets indicate the average over the Bose-Einstein  distribution 
for the momentum $\bm{q}$ of the pion with flavor $ad$ or $ac$.

The terms  $\omega_{adq}$, $\Delta_{ad}$, and $\omega_{acq}$ 
 in the denominator of the self-energy tensor in Eq.~\eqref{PiD*aij} are order $m_\pi$.
The heavy-meson expansion is obtained by expanding in powers of the small energies
$E - \varepsilon_{\ast a}$, $E - \varepsilon_{\ast c}$, $(\bm{p} \pm \bm{q})^2/(2M)$,
and $(\bm{p} \pm \bm{q})^2/(2M_*)$.
After averaging over the directions of the pion momentum, terms with odd powers of $\bm{q}$ vanish
and the self-energy tensor reduces to the diagonal tensor $\Pi_{\ast a}(E,p)\,  \delta^{mn}$.
The  LO term is a constant independent of $E$ and $p$:
\beq
\Pi_{\ast a}^\mathrm{(LO)} = 
-\frac{g_\pi^2}{6 f_\pi^2} \, \mathfrak{n}_\pi  \sum_d (2 - \delta _{ad} ) \,\Delta_{ad}\, \mathcal{F}_{ad}^\ast,
\label{PiD*a-LO}
\eeq
where $\mathcal{F}_{ad}^\ast$ is the complex conjugate of the thermal average defined in Eq.~\eqref{Fcd}.
The NLO term is a linear function of $E$ and $p^2$: 
\bqa
\Pi_{\ast a}^\mathrm{(NLO)}(E,p) &=&- \frac{g_\pi^2}{6 f_\pi^2}  \, \mathfrak{n}_\pi 
\left\{\sum_d (2 - \delta _{ad} ) 
\left[ \left(E - \varepsilon_{\ast a} - \frac{p^2}{2M}\right) \mathcal{G}_{1,ad}^\ast
 - \frac{1}{2M}  \mathcal{G}_{2,ad}^\ast \right] \right.
\nonumber\\
&&\hspace{0cm} \left. 
+ \sum_c (2 - \delta _{ac} ) \left[
2\left(E - \varepsilon_{\ast c} - \frac{p^2}{2M_\ast} \right)
\left \langle \frac{q^2}{\omega_{acq}^3} \right\rangle 
- \frac{1}{M_\ast} \left \langle \frac{q^4}{\omega_{acq}^3} \right\rangle \right]\right\},
\label{PiD*a-NLO}
\eqa
where $\mathcal{G}_{n,ad}^\ast$ is the complex conjugate of the thermal average defined in Eq.~\eqref{Gcd,n}.
We can read off the corrections $\delta  \varepsilon_{\ast a}$, $\delta Z_{\ast a}$, and $\zeta_{\ast a}$  
defined by the expansion in Eq.~\eqref{D*apropexpand}:
\begin{subequations}
\bqa
\delta  \varepsilon_{\ast a} &=& - \frac{g_\pi^2}{6f_\pi^2} \, \mathfrak{n}_\pi  
\left\{\sum_d (2 - \delta _{ad} )
\left[   \Delta_{ad}\, \mathcal{F}_{ad}^\ast - \frac{1}{2M}   \mathcal{G}_{2,ad}^\ast \right] \right.
\nonumber\\
&& \hspace{2cm} \left.-  \sum_c (2 - \delta _{ac} ) \left[ 2 (\varepsilon_{*c}- \varepsilon_{*a})
\left \langle \frac{q^2}{\omega_{acq}^3} \right\rangle 
+ \frac{1}{M_\ast} \left \langle \frac{q^4}{\omega_{acq}^3} \right\rangle \right] \right\},
\label{deltaepsilonD*a}
\\
\delta Z_{\ast a} &=& - \frac{g_\pi^2}{6 f_\pi^2} \, \mathfrak{n}_\pi  
\left\{ \sum_d (2 - \delta _{ad} )   \mathcal{G}_{1,ad}^\ast
 + 2  \sum_c (2 - \delta _{ac} ) \left \langle \frac{q^2}{\omega_{acq}^3} \right\rangle \right\},
\label{deltaZD*a}
\\
\zeta_{\ast a} &=&  \frac{g_\pi^2 \, \Delta}{6 f_\pi^2 M}  \, \mathfrak{n}_\pi \,
\sum_d(2 - \delta _{ad} ) \, \mathcal{G}_{1,ad}^\ast.
\label{deltazetaD*a}
\eqa
\label{deltaD*a}%
\end{subequations}
For a pion gas at $T_\mathrm{kf} = 115$\,MeV, the thermal rest energies for $D^{\ast +}$ and $D^{\ast 0}$ 
are  $\delta  \varepsilon_{\ast +} = (-0.21 - 0.013 \,i)$\,MeV 
and $\delta  \varepsilon_{\ast 0} = (-0.16 - 0.006 \,i)$\,MeV. 
Their real parts are the thermal mass shifts.
The thermal widths are their imaginary parts multiplied by $-2$: 
$\delta \Gamma_{\ast b} = -2\, \mathrm{Im}[\varepsilon_{\ast b}]$.
The changes in the residue factors are $\delta Z_{\ast +}  = -0.012 - 0.004\, i$
and $\delta Z_{\ast 0}  =-0.011 - 0.001\, i$.

\subsection{Charm-meson Rest Energies}
\label{sec:EnergyShift}

The complex thermal rest energies of the charm mesons
in Eqs.~\eqref{deltaepsilonDb} and \eqref{deltaepsilonD*a} 
can be  expressed as double expansions in isospin splittings divided by $m_\pi$ and in
pion/charm-meson mass ratios using results given in Appendix~\ref{app:PionIntegral}.
The leading terms in the thermal mass shifts for the $D$ and $D^*$ do not depend on their flavors:
\begin{subequations}
\bqa
\mathrm{Re}[\delta \varepsilon_b] &\approx& 
\frac{3g_\pi^2 m_\pi}{2f_\pi^2} \,  \mathfrak{n}_\pi \, \left\langle \frac{1}{\omega_q} \right\rangle,
\label{Re[deltaepsilonb]}
\\
\mathrm{Re}[\delta \varepsilon_{\ast a}] &\approx&
-\frac{g_\pi^2 m_\pi}{2f_\pi^2} \,  \mathfrak{n}_\pi \, \left\langle \frac{1}{\omega_q} \right\rangle.
\label{Re[deltaepsilon*c]}
\eqa
\label{Re[deltaepsilon]}%
\end{subequations}
For a pion gas at $T_\mathrm{kf} = 115$\,MeV,
these thermal mass shifts are $\mathrm{Re}[\delta \varepsilon_b]=+1.26$\,MeV
and $\mathrm{Re}[\delta \varepsilon_{\ast a}]=-0.42$\,MeV. 
The more accurate results for $D^+$ and $D^0$ from  Eq.~\eqref{deltaepsilonDb} 
differ by factors of 1.27 and 1.41, respectively.
The more accurate results for $D^{\ast +}$ and $D^{\ast 0}$ from  Eq.~\eqref{deltaepsilonD*a} 
differ by factors of 0.51 and 0.38, respectively.
The spin-weighted average of the simple thermal mass shifts in Eq.~\eqref{Re[deltaepsilon]} is zero.
At $T_\mathrm{kf} = 115$\,MeV, the spin-weighted average 
of the more accurate  thermal mass shifts for the four charm mesons  is +0.28\,MeV.  

\begin{figure}[t]
\includegraphics[width=0.48\textwidth]{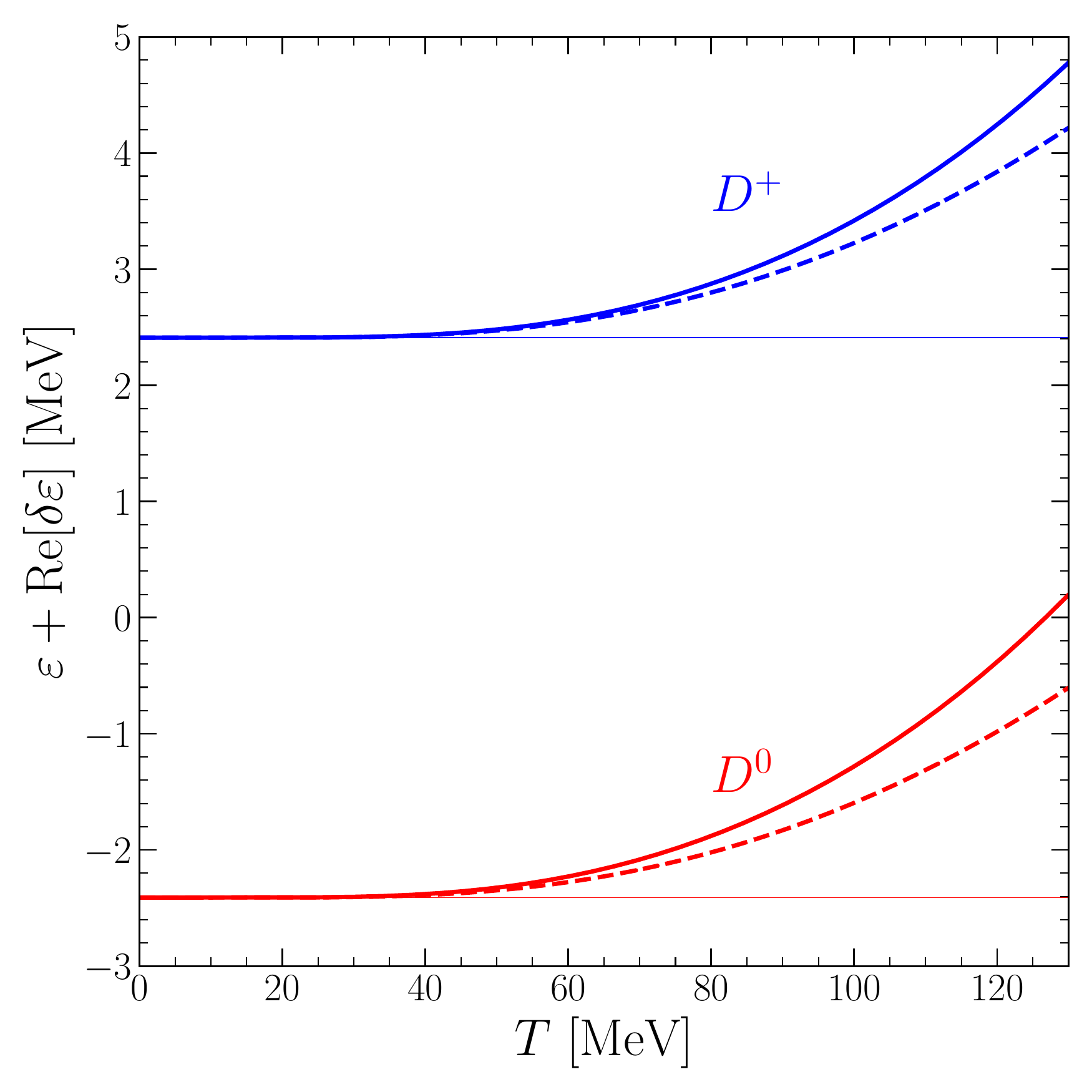} ~
\includegraphics[width=0.48\textwidth]{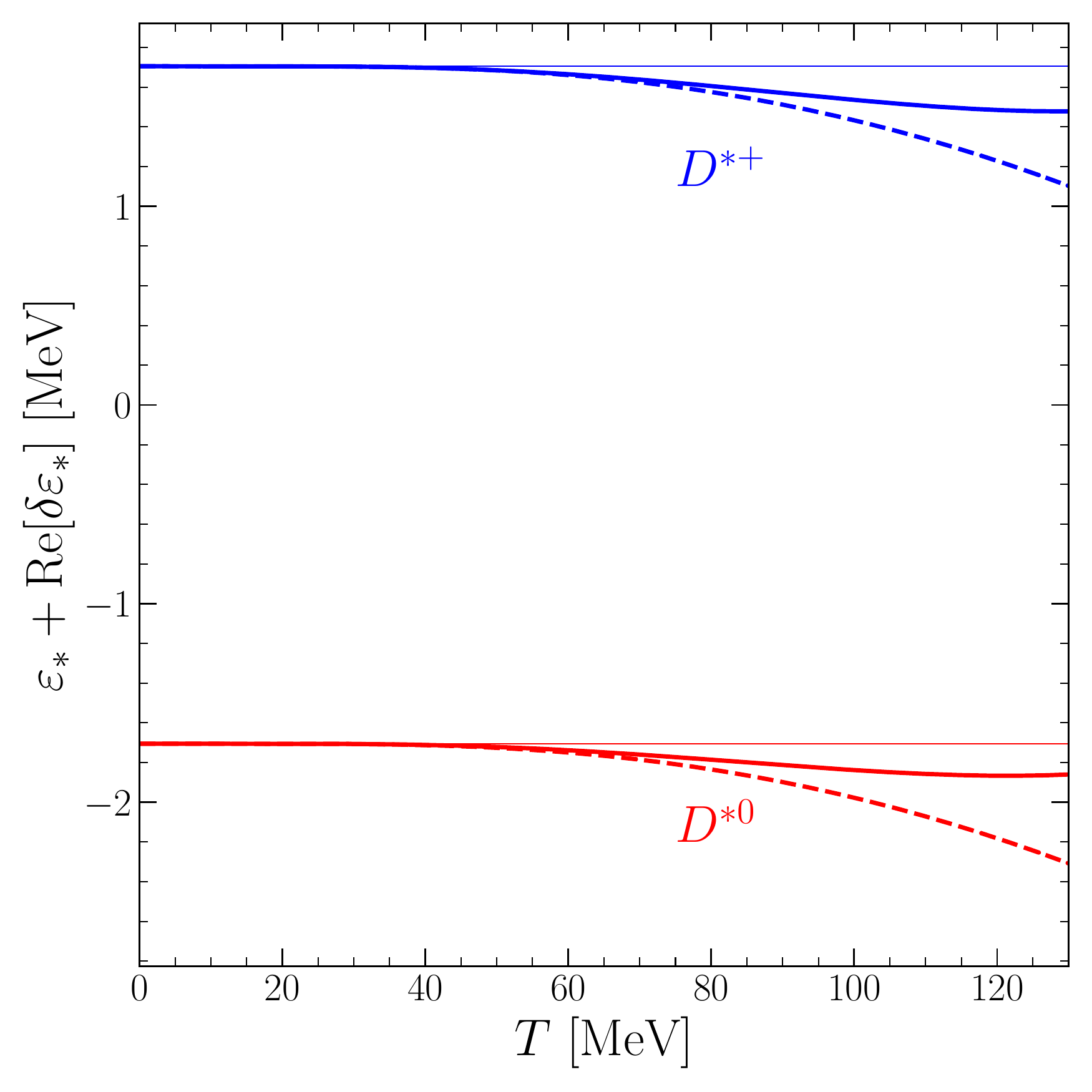}
\caption{
Real rest energies for $D^+$ and $D^0$ (left panel: blue upper and red lower
curve) and for $D^{\ast +}$ and $D^{\ast 0}$  (right panel: blue upper and red lower curve)
in a pion gas as functions of the temperature $T$.
The solid curve is the sum of the rest energy $\varepsilon$ at $T=0$ 
and the thermal mass shift $\mathrm{Re}[\delta \varepsilon]$.
The dashed curve uses the simple approximation for the thermal mass  shift in Eqs.~\eqref{Re[deltaepsilon]}. 
The horizontal line is the rest energy at $T=0$.
}
\label{fig:DD*energyshift}
\end{figure}

The temperature dependence of the real parts of the charm-meson rest energies is illustrated in Fig.~\ref{fig:DD*energyshift}.  
The range of the temperature $T$ is only up to 130\,MeV, 
which may already be beyond the range of validity of our results. 
A solid curve  is the sum of the rest energy at $T=0$ and the thermal mass shift.
The thermal mass shifts are smaller than the isospin splittings in this temperature range.
The thermal mass shifts for $D^+$ and $D^0$ increase monotonically as $T$ increases.
The  thermal mass shifts for $D^{*+}$ and $D^{*0}$ have local minima near 130 and 120 MeV, respectively.
A dashed curve in Fig.~\ref{fig:DD*energyshift} uses the approximation for the  thermal mass shift in Eqs.~\eqref{Re[deltaepsilon]}.
The approximations for $D$ and $D^\ast$ in Eqs.~\eqref{Re[deltaepsilonb]} and \eqref{Re[deltaepsilon*c]}
match well with the more accurate results in Eqs.~\eqref{deltaepsilonDb} and \eqref{deltaepsilonD*a}
at sufficiently low temperatures. 
The approximation for $D$ has the same qualitative behavior as the more accurate result at higher temperatures.
The approximation for $D^\ast$ has the same qualitative behavior as the more accurate result 
only at temperatures lower than about 100\,MeV.  
The difference at higher temperatures comes from the increasingly significant contributions 
from the thermal averages $\mathcal{G}_{2,ad}^*$ and $\left\langle q^4/\omega_{acq}^3\right\rangle$ in Eq.~\eqref{deltaepsilonD*a} at higher temperatures.

The imaginary parts of the charm-meson  thermal rest energies
in Eqs.~\eqref{deltaepsilonDb} and \eqref{deltaepsilonD*a} can be evaluated analytically. 
The leading terms in their expansions in isospin splittings divided by $m_\pi$ 
and in pion/charm-meson mass ratios can be expressed in terms of the partial decay rates for $D^\ast \to D \pi$:
\bea
\Gamma[D^{*c} \to D^d \pi ]  = 
\frac{g_{\pi}^{2}}{12 \pi f_{\pi}^{2}}\left(2-\delta_{cd}\right)
\left(\Delta_{cd}^{2}-m_{\pi cd}^{2}\right)^{3 / 2} \,\theta(\Delta_{cd}-m_{\pi cd}).
\label{eq:Gamma-DtcDapi}
\eea
The leading terms in the  thermal widths of the charm mesons are
\begin{subequations}
\bqa
\delta \Gamma_b &\approx&
3\,   \mathfrak{f}_\pi(m_\pi)\,   \sum_c  \Gamma[D^{*c} \to D^b \pi ] ,
\label{Im[deltaepsilonb]}
\\
\delta \Gamma_{*a} &\approx& 
 \mathfrak{f}_\pi(m_\pi)\,  \sum_d  \Gamma[D^{*a} \to D^d \pi ] .
\label{Im[deltaepsilon*a]}
\eqa
\label{Im[deltaepsilon]}%
\end{subequations}
The isospin average for $D^\ast$ is 1/3 that for $D$.
The experimental values of the three nonzero charm-meson pionic decay rates at zero temperature are
$\Gamma[D^{*+}\to D^0\pi^+] = 56.5$\,keV,
$\Gamma[D^{*0}\to D^0\pi^0] = 35.8$\,keV, and
$\Gamma[D^{*+}\to D^+\pi^0] = 25.6$\,keV. 
(Recall that $\Gamma[D^{*+}\to D^0\pi^+]$ was used to determine $g_\pi$.)
For a pion gas at $T_\mathrm{kf} = 115$\,MeV, 
the value of the Bose-Einstein distribution in Eqs.~\eqref{Im[deltaepsilon]} is $\mathfrak{f}_\pi(m_\pi) = 0.431$.
The thermal widths of  $D^+$ and $D^0$ from Eq.~\eqref{Im[deltaepsilonb]}
are $\delta \Gamma_+ =33$\,keV and $\delta \Gamma_0 =119$\,keV. 
The more accurate results from  Eq.~\eqref{deltaepsilonDb} differ by +10\% and +8\%.  
The thermal widths of  $D^{\ast +}$ and $D^{\ast 0}$ from Eq.~\eqref{Im[deltaepsilon*a]}
are $\delta \Gamma_{\ast +} =35$\,keV and $\delta \Gamma_{\ast 0} =15$\,keV. 
The more accurate results from  Eq.~\eqref{deltaepsilonD*a} differ by $-25\%$ and $-24\%$.

\begin{figure}[t]
\includegraphics[width=0.48\textwidth]{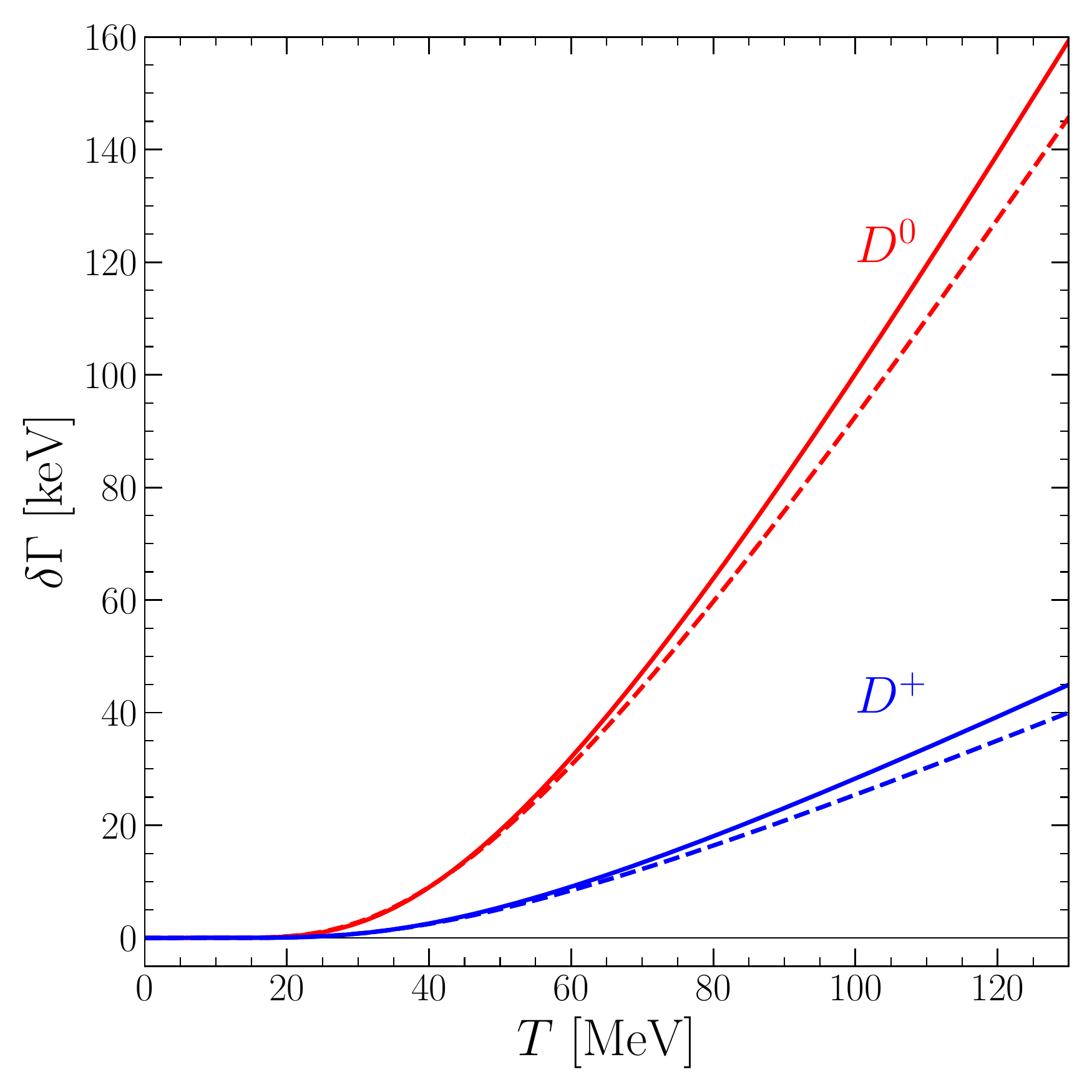} ~
\includegraphics[width=0.48\textwidth]{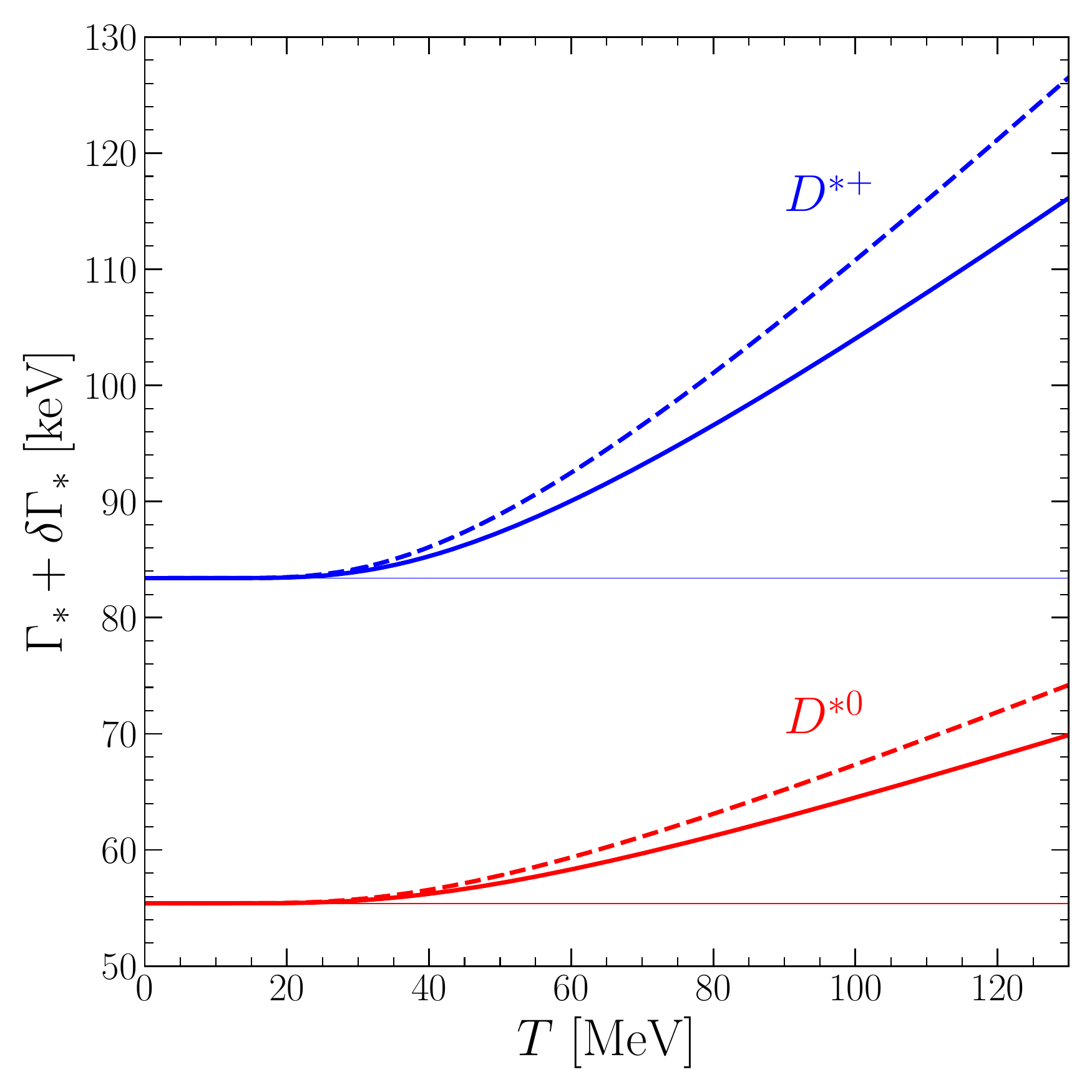}
\caption{
Widths for $D^+$ and $D^0$ (left panel: blue lower and red upper curve) and for
$D^{\ast +}$ and $D^{\ast 0}$  (right panel: blue upper and red lower curve)
in a pion gas as functions of the temperature $T$.
A solid curve is the sum of the decay width $\Gamma$ at $T=0$ 
and the thermal width $\delta \Gamma$.
A dashed curve uses the simple approximation for the thermal width in Eqs.~\eqref{Im[deltaepsilon]}. 
A horizontal line is the decay width at $T=0$.
}
\label{fig:DD*width}
\end{figure}

The widths for the charm mesons as functions of 
the temperature $T$ are shown in Fig.~\ref{fig:DD*width}.
For a pseudoscalar charm meson $D^b$, the solid curve is the thermal width $\delta \Gamma_b$.
For a vector charm meson $D^{\ast a}$, the solid curve is
 the sum  $\Gamma_{\ast a} + \delta \Gamma_{\ast a}$ of the decay width and the thermal width.
The thermal widths all increase monotonically with $T$.
A dashed curve  in Fig.~\ref{fig:DD*width} uses the approximation for the  thermal width in Eqs.~\eqref{Im[deltaepsilon]}.
The approximations for $D$ and $D^\ast$ in Eqs.~\eqref{Im[deltaepsilonb]} and \eqref{Im[deltaepsilon*a]}
match well with the more accurate results in Eqs.~\eqref{deltaepsilonDb} and \eqref{deltaepsilonD*a}
at sufficiently low temperature.
The approximations have the same qualitative behavior as the more accurate results at higher temperatures.

\subsection{Comparison with Previous Work}
\label{sec:DD*compare}

There have been several previous calculations of the  thermal energies of charm mesons in a pion gas  \cite{Fuchs:2004fh,He:2011yi,Cleven:2017fun,Montana:2020lfi,Montana:2020vjg}. 
The temperatures ranged from 0 to beyond the hadronization temperature, which is near 150\,MeV. 
Our results are not applicable at temperatures as high as 150\,MeV.
We will therefore only  compare our results with previous calculations at 100\,MeV. 
The comparisons are presented in Table~\ref{tab:charm-meson comparison}.

The complex thermal energies of the charm mesons $D$ and $D^\ast$ in a pion gas
have been calculated by Fuchs {\it et al.}\ using a resonance approach \cite{Fuchs:2004fh}.
Their $\pi D^{(\ast)}$ elastic scattering amplitudes 
are the sums of relativistic Breit-Wigner amplitudes corresponding to the 
known S-wave, P-wave, and D-wave $\pi D^{(\ast)}$ resonances.
These amplitudes are not consistent with the low-energy constraints from chiral symmetry.
The thermal self energies from pion forward scattering 
were obtained by integrating the  $\pi D^{(\ast)}$ scattering amplitudes over the pion momentum distribution.
The results of Fuchs {\it et al.}\  for the $D$ and $D^*$ thermal mass shifts and the $D$ and $D^*$ thermal widths at $T=100$~MeV are given in Table~\ref{tab:charm-meson comparison}.
Similar results for the $D$ thermal width were obtained 
by He {\it et al.}\  using a resonance approach \cite{He:2011yi}.

The  thermal widths of $D$ and $D^\ast$ in a pion gas
have been calculated by Cleven {\it et al.}\  using a self-consistent unitarized approach \cite{Cleven:2017fun}.
Their tree-level $\pi D^{(\ast)}$ amplitudes are vertices from an effective lagrangian
 for pseudoscalar and vector mesons with an approximate $SU(4)$ symmetry.
They did not include diagrams with intermediate $D$ or $D^\ast$ propagators,
like those in Figs.~\ref{fig:Dselfenergy} and \ref{fig:D*selfenergy}.
Their unitarized $\pi D^{(\ast)}$ amplitudes are the analytic solutions 
of the corresponding regularized Lippmann-Schwinger equations.
A thermal  $D^{(\ast)}$ self energy can be obtained numerically as a function of the 3-momentum  
by integrating the  $\pi D^{(\ast)}$ amplitude over the loop momentum and over the pion momentum distribution.
A Lippmann-Schwinger integral equation for the thermal $\pi D^{(\ast)}$ amplitude 
can be obtained by inserting the thermal  $D^{(\ast)}$ self energy into the $D^{(\ast)}$ propagator in the loop.
Self-consistent solutions  for the thermal  $D^{(\ast)}$ self energy and the thermal $\pi D^{(\ast)}$ amplitude 
were obtained  by iterating the equations several times numerically. 
Their results for the thermal  mass shifts of  $D$ and $D^\ast$ were little more than numerical noise.
The results of Clevens {\it et al.}\ for the $D$ and $D^*$ thermal widths at $T=100$~MeV are given in
Table~\ref{tab:charm-meson comparison}.

The complex thermal energies of $D$ in a pion gas and in a gas of $\pi$, $K$ and $\eta$ mesons
have been calculated by Monta\~na {\it et al.}\  using a different self-consistent unitarized approach  \cite{Montana:2020lfi}.
Their low-energy $\pi D$ scattering amplitude is  the sum of the contact vertex in Fig.~\ref{fig:Dselfenergy}
and the next-to-leading order vertices in heavy-hadron $\chi$EFT.
They did not include the diagram with  the intermediate $D^\ast$ propagator in Fig.~\ref{fig:Dselfenergy}. 
Their unitarized amplitude is the analytic solution of the corresponding regularized Lippmann-Schwinger equation.
The unitarization reduces the accuracy to leading order in the chiral expansion.
Self-consistent solutions to the integral for the thermal $D$ self energy
and the  Lippmann-Schwinger integral equation for the thermal $\pi D$ amplitude were obtained numerically.
The results were extended to other charm mesons including $D^*$ in Ref.~\cite{Montana:2020vjg}.
The results of Montana {\it et al.}\ for the $D$ and $D^*$ thermal mass shifts and the $D$ and $D^*$ thermal widths at $T=100$~MeV are given in Table~\ref{tab:charm-meson comparison}.

\begin{table}[t]
    \centering
    \begin{tabular}{cccccc}
        & &  Re$[\delta \varepsilon]$ & Re$[\delta \varepsilon_*]$ & $\delta \Gamma$ & $\delta \Gamma_*$ \\ 
         \toprule
        Fuchs \textit{et al.} \cite{Fuchs:2004fh} & & $-7.0$ & $-6.0$ & 8 & 12 \\ 
        Cleven \textit{et al.} \cite{Cleven:2017fun} & & &  & 15 & 10 \\ 
        Monta\~na \textit{et al.}  \cite{Montana:2020lfi,Montana:2020vjg}& & $-13$ & $-12$ & 17 & 17 \\ 
        \hline
        This work & & 1.07 & $-0.15$ & 0.064 & 0.015 \\
       \bottomrule
       \hline
    \end{tabular}
    \caption{The flavor-averaged thermal mass shifts Re$[\delta\varepsilon_{(\ast)}]$ and thermal widths $\delta\Gamma_{(\ast)}$ (in MeV)
    for charm mesons at $T=100$ MeV. 
    }\label{tab:charm-meson comparison}
\end{table}
The thermal mass shifts and thermal widths for $D$ and $D^\ast$ at $T=100$~MeV calculated in Refs.~\cite{Fuchs:2004fh,He:2011yi,Cleven:2017fun,Montana:2020lfi,Montana:2020vjg} 
differ at most by about a factor of 2.
Our results in Table~\ref{tab:charm-meson comparison} for
the flavor averages at $T=100$\,MeV of the thermal mass shifts for $D$ from Eq.~\eqref{deltaepsilonDb} 
and for $D^\ast$ from Eq.~\eqref{deltaepsilonD*a} are $+1.07$\,MeV and  $-0.15$\,MeV, 
which have opposite signs.
The  thermal mass shifts  for $D$ and $D^\ast$  in Refs.~\cite{Montana:2020lfi,Montana:2020vjg} 
are both negative, approximately equal, and
larger than our results by about one and two orders of magnitude, respectively.
Our results in Table~\ref{tab:charm-meson comparison} for the flavor averages at $T=100$\,MeV of the thermal widths for $D$ from Eq.~\eqref{deltaepsilonDb} 
and for $D^\ast$ from Eq.~\eqref{deltaepsilonD*a} are 64\,keV and 15\,keV. 
The thermal widths for $D$ and $D^\ast$ in Refs.~\cite{Montana:2020lfi,Montana:2020vjg}
are approximately equal and larger than our results by about 3 orders of magnitude.

The orders-of-magnitude discrepancies between our results
and those in Refs.~\cite{Montana:2020lfi,Montana:2020vjg} are a serious issue.
We cannot exclude the possibility that higher-order terms  in the chiral expansions of the thermal 
rest energies for charm mesons are much larger than the leading-order terms,
even at temperatures as low as 100\,MeV.
In this case, our thermal masses and thermal widths  for charm mesons could be much too small.
One possible source of large corrections is from S-wave pion interactions,
since the contributions from total isospin  $\tfrac12$ and $\tfrac32$ cancel at leading order.  
All previous calculations of S-wave $\pi D$ scattering lengths at higher order in the chiral expansion,
including those obtained using a chiral unitarization prescription, 
imply that the S-wave contribution to the $D$ thermal mass shift is smaller 
than the $I=\tfrac12$ contribution at leading order,
which can be obtained from Eq.~\eqref{PiS} by replacing the factor (1-1) by 1.
For a pion gas at $T_\mathrm{kf} = 100$~MeV, the $I = \tfrac12$ contribution is 4.3~MeV, 
which is about 4 times larger than the flavor-averaged thermal mass shift from P-wave interactions.
The $D$ mass shift in Ref.~\cite{Montana:2020lfi} is larger by another factor of 3, 
and it has the opposite sign.
So this possible source of large corrections seems to be insufficient.
Another possible explanation for the large discrepancies between our results
and those in Refs.~\cite{Montana:2020lfi,Montana:2020vjg}  is that they arise from
the iteration of the integral equation used to obtain self-consistent solutions.
This integral equation was obtained by generalizing the integral equation for a chiral unitarization prescription to nonzero temperature.
The chiral unitarization prescription is just a model, and the proposed generalization to finite temperature
may give thermal mass shifts and thermal widths for charm mesons 
that are much too large at low temperatures.
In Refs.~\cite{Montana:2020lfi,Montana:2020vjg},
the S-wave contributions to $\pi D$ scattering were taken into account, but the P-wave contributions were ignored. 
This is appropriate only at temperatures much smaller than $m_\pi$.  
The effects of the P-wave contributions could be  significant at $T=100$\,MeV.

The first lattice QCD  calculations of charm meson masses as functions of the temperature 
were presented in Ref.~\cite{Aarts:2022krz}.
The lattice QCD simulations were carried out with 3 flavors of dynamical light quarks and a pion mass of  239\,MeV.
The results have not been extrapolated to the physical pion mass.
The masses of $D$ and $D^\ast$ were calculated at 6 temperatures ranging from 47 to 169\,MeV.
At the 4 lowest temperatures, which range from 47~MeV to 127~MeV,
the thermal shifts in the masses  of $D$ and $D^\ast$
are consistent with zero to within the errors, which are about 6\,MeV.
The lowest temperature where the thermal mass shifts are significant is 152\,MeV,
where the shift is $-20 \pm 7$\,MeV for $D$ and  $-43 \pm 10$\,MeV for $D^\ast$.
The $D$ thermal mass shift at 150 MeV in Refs.~\cite{Montana:2020lfi,Montana:2020vjg}
is larger by a factor of 2, which is about 3 standard deviations, 
while the $D^\ast$ thermal mass shift is compatible with Ref.~\cite{Aarts:2022krz}.
At a temperature of 109~MeV, the thermal shifts in Ref.~\cite{Aarts:2022krz}
are $0\pm 6$~MeV for $D$ and $+4 \pm 6$~MeV for $D^\ast$.
The $D$ and $D^\ast$ thermal shifts at 100 MeV in Refs.~\cite{Montana:2020lfi,Montana:2020vjg}
are outside the error bar by 2.2 and 2.7 standard deviations, respectively.
Thus lattice QCD is beginning to provide useful results for the thermal shifts in charm-meson masses.


\section{Charm-meson Pair Self Energy}
\label{sec:CharmPair}

In this Section, we calculate the $D^\ast D$ self energy in the pion gas
to NLO in the heavy-meson expansion.
For simplicity, we calculate it only for charm-meson pairs with kinetic energies of order $m_\pi^2/M$.
The self-energy is complicated by terms that diverge at the charm-meson-pair threshold
and by terms that correspond to ultraviolet divergences in ZREFT.

\subsection{Leading Order}
\label{PiX-LO}

\begin{figure}[t]
\includegraphics[width=0.99\textwidth]{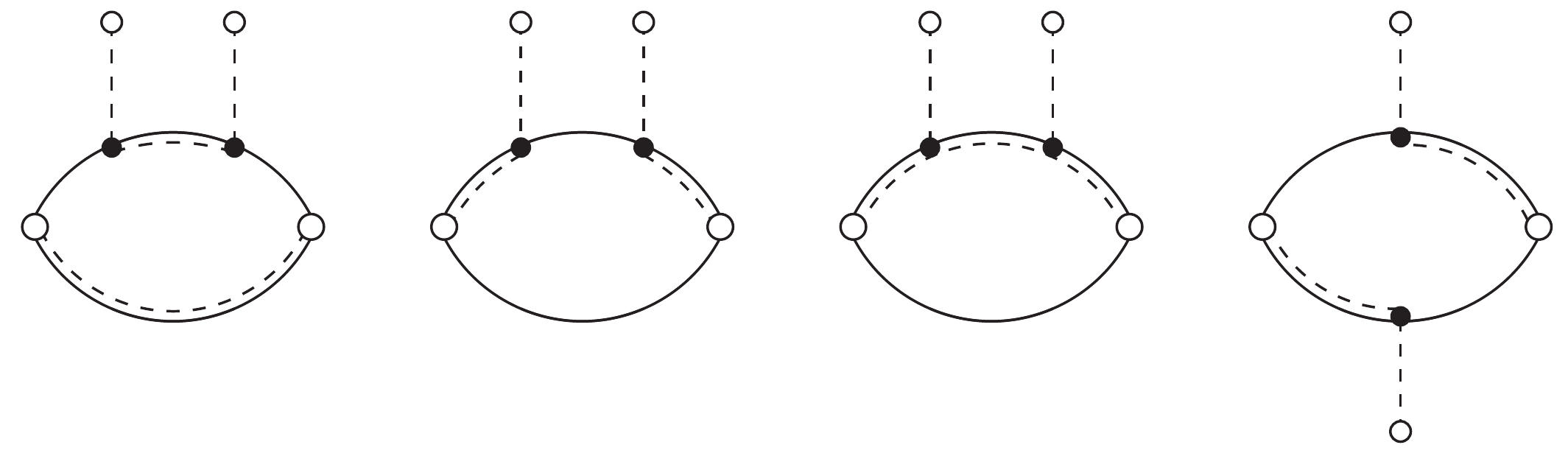}
\caption{
One-loop diagrams for the $D^\ast D$ self energy from pion forward scattering.
The diagrams are summed over the two directions for the routing of the pion momentum.
The contribution to $i (\mu/2\pi)\Sigma$ is the sum of the diagrams with pion legs amputated, 
weighted by $\mathfrak{f}_\pi(\omega_q)/(2\omega_q)$,
integrated over $\bm{q}$, and summed over the pion flavor $i$.
The diagrams with a $\pi D$ contact vertex or a $\pi D^\ast$ contact vertex 
are not shown, because each diagram cancels upon summing over the pion flavor.
}
\label{fig:Xselfenergy}
\end{figure}

At leading order in the pion interactions, the thermal contributions to the $D^\ast D$ self-energy come
from pion forward scattering. 
Four of the corresponding diagrams for the $D^\ast D$ self energy are shown in Fig.~\ref{fig:Xselfenergy}.
There are two additional diagrams with $\pi D$  or $\pi D^\ast$ contact vertices.
They are not shown, because they cancel upon summing over the pion flavor.
Each pion-forward-scattering diagram is the sum of two terms:
an on-shell pion  emerges from one of the two open circles with momentum $\bm{q}$ and flavor $i$
and it scatters into the other open circle with the same momentum and flavor.
The amplitudes must be added coherently by multiplying them by $\mathfrak{f}_\pi(\omega_q)/(2\omega_q)$,
integrating over $\bm{q}$  with measure $d^3q/(2\pi)^3$, and summing over the flavors $i$.

The self energy diagrams in Fig.~\ref{fig:Xselfenergy} are 
functions of the total energy $E$ and total momentum $\bm{P}$ of the $D^\ast D$ pair. 
They are integrated over the pion momentum $\bm{q}$,
the loop momentum $\bm{k}$, and the loop energy $\omega$.
The integral over $\omega$ can be evaluated by contours.
In the first three diagrams in Fig.~\ref{fig:Xselfenergy},
we use the pole of the single propagator connecting the $D^\ast D$ pair vertices.
The large pion momentum of order $m_\pi$ 
must then flow through the single propagator connecting the pion vertices.
The other two propagators have identical denominators with small energies of order $m_\pi^2/M$.
In the fourth diagram in Fig.~\ref{fig:Xselfenergy},
we integrate over $\omega$ using the poles of the two adjacent propagators on the lower side of the loop.
The large pion momentum of order $m_\pi$ must flow through the two propagators 
on either the right side or the left side of the loop.
The remaining propagator on the upper left or upper right of the loop 
has a denominator with a small energy of order $m_\pi^2/M$.

In a charm-meson propagator carrying the large momentum $\bm{q}$, 
the large energies in the denominator are in terms of the form $\omega_{cdq}$ and $\Delta_{cd}$,
which are order $m_\pi$.
The heavy-meson expansion is obtained by expanding in powers of small energies of order $m_\pi^2/M$,
which include the energy difference $E-\varepsilon_{\ast a} - \varepsilon_b$,
the kinetic energies of the charm mesons, and isospin splittings. 
Only the first  three diagrams in Fig.~\ref{fig:Xselfenergy} have contributions at LO in that expansion.
All three diagrams factor into an integral over $\bm{q}$ and the same convergent integral over $\bm{k}$,
which is given in Eq.~\eqref{softint} of Appendix~\ref{app:MomIntegral}. 
The LO contributions from the third diagram cancel upon summing over the two directions for the routing of the pion 4-momentum,  so it is actually NLO.  In the first two diagrams,
the integrand of the integral over $\bm{q}$ can be simplified by summing 
over the two directions  for the routing of the pion 4-momentum. 
The contributions to the $D^\ast D$ self energy at  LO
from the first two diagrams in Fig.~\ref{fig:Xselfenergy} are 
\begin{subequations}
\bqa
\Sigma_1^\mathrm{(LO)} &=&
\frac{g_\pi^2\mu}{2 f_\pi^2} \,  \mathfrak{n}_\pi \,  \frac{1}{S_0(E_\mathrm{cm}) }
\, \sum_c (2-\delta_{cb}) \,\Delta_{cb}\, \mathcal{F}_{cb}  ,
\\
\Sigma_2^\mathrm{(LO)}&=& 
-\frac{g_\pi^2\mu}{6 f_\pi^2} \, \mathfrak{n}_\pi  \, \frac{1}{S_0(E_\mathrm{cm}) }
 \sum_d  (2-\delta_{ad})  \Delta_{ad}\,  \mathcal{F}_{ad}^\ast ,
\eqa
\label{SigmaXLO}%
\end{subequations}
where $\mathcal{F}_{cd}$ is the thermal average defined in Eq.~\eqref{Fcd} and $S_0( E_\mathrm{cm} )$ is 
the function of the center-of-mass energy defined in Eq.~\eqref{S0-Ecm}.

The LO contribution to the $D^\ast D$ self energy is the sum of Eqs.~\eqref{SigmaXLO}.
It can be expressed as $\mu(\delta \varepsilon_{*a} + \delta \varepsilon_b)/S_0(E_\mathrm{cm})$,
where $\delta \varepsilon_b$ and $\delta \varepsilon_{*a}$ 
are the thermal rest energies of the charm mesons at LO, which are given by the 
$\mathcal{F}_{cb}$ term in Eq.~\eqref{deltaepsilonDb}
and by the $\mathcal{F}_{ad}^\ast$ term in Eq.~\eqref{deltaepsilonD*a}.
This agrees with the LO term in the expansion of $S_1(E_\mathrm{cm},p)$ around $S_0(E_\mathrm{cm})$ 
in Eq.~\eqref{S1-S0}.  Thus the $D^\ast D$ self energy at LO can be completely absorbed into the 
complex thermal rest energies of the  charm-meson constituents.

\subsection{Next-to-Leading Order from One-loop Diagrams}
\label{PiX-NLO}

The $D^\ast D$ self energy has NLO contributions from all four diagrams in Fig.~\ref{fig:Xselfenergy}.
The NLO contributions from the first three diagrams are obtained by expanding the propagator
carrying the large momentum $\bm{q}$ to  first order in the small energies.
The resulting integrals over the loop momentum can be reduced to the forms found in
Appendix~\ref{app:MomIntegral}: Eq.~\eqref{softint},
which gives a factor of $1/S_0(E_\mathrm{cm})$,
and Eqs.~\eqref{int2-k} and \eqref{int2-Pmk},  which are ultraviolet divergent.
The  NLO contribution from the fourth diagram in Fig.~\ref{fig:Xselfenergy}
can be obtained by setting the small energies in the two propagators carrying the large momentum
 equal to 0.
The resulting integral over the loop momentum  reduces to Eq.~\eqref{int1}, which is ultraviolet divergent.
The NLO terms in the four self-energy diagrams are
\begin{subequations}
\bqa
\Sigma_1^\mathrm{(NLO)} &=&
\frac{g_\pi^2 M}{f_\pi^2 M_X} \,  \mathfrak{n}_\pi \,\Bigg\{ 
 \left[ \Lambda - \left(1 - \frac{\Delta}{4 M} \right)S_0(E_\mathrm{cm})  
- \frac{M \Delta}{4 M_X^2}  \frac{P^2 }{S_0(E_\mathrm{cm})}\right] 
 \sum_c (2-\delta_{cb})  \mathcal{G}_{1,cb}
\nonumber\\
&& \hspace{2cm}
+ \frac{1}{4\, S_0(E_\mathrm{cm})}
 \sum_c (2-\delta_{cb})  \mathcal{G}_{2,cb} \Bigg\} ,
\label{Sigma1NLO}
\\
\Sigma_2^\mathrm{(NLO)}  &=&
\frac{g_\pi^2 M_\ast}{3f_\pi^2 M_X} \,  \mathfrak{n}_\pi  \,\Bigg\{ 
 \left[ \Lambda - \left(1+\frac{\Delta}{4M_\ast} \right)S_0(E_\mathrm{cm})  
+ \frac{M_\ast \Delta}{4 M_X^2}  \frac{P^2 }{S_0(E_\mathrm{cm})}\right]
\sum_d (2-\delta_{ad}) \mathcal{G}_{1,ad}^*
\nonumber\\
&& \hspace{2cm}
+ \frac{1}{4\,S_0(E_\mathrm{cm})}
\sum_d (2-\delta_{ad}) \mathcal{G}_{2,ad}^* \Bigg\} ,
\label{Sigma2NLO}
\\
\Sigma_3^\mathrm{(NLO)}  &=&
\frac{g_\pi^2}{3f_\pi^2} \,  \mathfrak{n}_\pi \,  \Bigg\{
 \big[ \Lambda - S_0(E_\mathrm{cm}) \big] 
  \sum_c (2-\delta_{ac}) \left\langle  \frac{q^2}{\omega_{acq}^3} \right\rangle 
\nonumber\\
&& \hspace{1cm}
+ \frac{1}{S_0(E_\mathrm{cm}) }
  \sum_c (2-\delta_{ac}) \left[ \mu(\varepsilon_{*c}-\varepsilon_{*a})
\left\langle  \frac{q^2 }{\omega_{acq}^3} \right\rangle  
+ \frac{M}{2M_X} \left\langle  \frac{q^4}{\omega_{acq}^3}\right\rangle  \right] \Bigg\} ,
\label{Sigma3NLO}
\\
\Sigma_4^\mathrm{(NLO)} &=&
 \frac{g_\pi^2}{3 f_\pi^2} \,  \mathfrak{n}_\pi \, \big[ \Lambda - S_0(E_\mathrm{cm})  \big]\,
  (2-\delta_{ab})\, \mathcal{H}_{ab},
\label{Sigma4NLO}
\eqa
\label{SigmaNLO}%
\end{subequations}
where $\Lambda$ is the ultraviolet cutoff, $\Delta = M_\ast - M$, and 
$\mathcal{G}_{n,cd}$ and $\mathcal{H}_{ab}$ are the thermal averages defined in Eqs.~\eqref{Gcd,n} and \eqref{Hcd}.

There are terms in the $D^\ast D$ self energy at NLO that diverge
at the branch point of $S_0(E_\mathrm{cm})$.
We compare them with the terms expected from the expansion of the square-root function 
$S_1(E_\mathrm{cm},P)$ in Eq.~\eqref{S1-S0}.
The sum of the singular terms proportional to $1/S_0(E_\mathrm{cm})$ in Eq.~\eqref{SigmaNLO}
matches the contribution to the term
$\mu (\delta\varepsilon_\ast + \delta\varepsilon) /S_0(E_\mathrm{cm})$
in Eq.~\eqref{S1-S0} from the NLO thermal $D^b$ and $D^{\ast a}$ rest energies
$\delta \varepsilon_b$ and $\delta \varepsilon_{\ast b}$
in Eqs.~\eqref{deltaepsilonDb} and \eqref{deltaepsilonD*a}. 
The sum of the singular terms proportional to $P^2/S_0(E_\mathrm{cm})$ in Eq.~\eqref{SigmaNLO}
can be expressed in the form of the last term in Eq.~\eqref{S1-S0}  with the coefficient
\beq
\zeta_X  =
- \frac{g_\pi^2\Delta }{2f_\pi^2 M_X} \,  \mathfrak{n}_\pi 
\left[ \frac{M}{M_\ast} \sum_c (2-\delta_{cb}) \mathcal{G}_{1,cb}
-\frac{M_\ast}{3M} \sum_d (2-\delta_{ad}) \mathcal{G}_{1,ad}^*   \right].
\label{zetaX-NLO}
\eeq
An expression for $\zeta_X$ that takes into account propagator corrections for the constituents
was deduced in Eq.~\eqref{zetaX}.
Its expansion to first order in $\zeta$ and $\zeta_\ast$ gives
\beq
\zeta_X  =  (M/M_X)\zeta_b + (M_\ast/M_X)  \zeta_{\ast a}.
\label{zetaX-1}
\eeq
Upon inserting the NLO expressions for $\zeta_b$ and $\zeta_{\ast a}$ in Eqs.~\eqref{deltazetaDb} and \eqref{deltazetaD*a}, we reproduce the expression for $\zeta_X$ in Eq.~\eqref{zetaX-NLO}.
Thus all  the terms in the $D^\ast D$ self energy at NLO
that diverge at the branch point of $S_0(E_\mathrm{cm})$
can be absorbed into the square-root function $S_1(E_\mathrm{cm},P)$ defined in Eq.~\eqref{S1-E}.

\subsection{Next-to-Leading Order from a Two-loop Diagram}
\label{Sigma2loop-NLO}

\begin{figure}[t]
\includegraphics[width=0.4\textwidth]{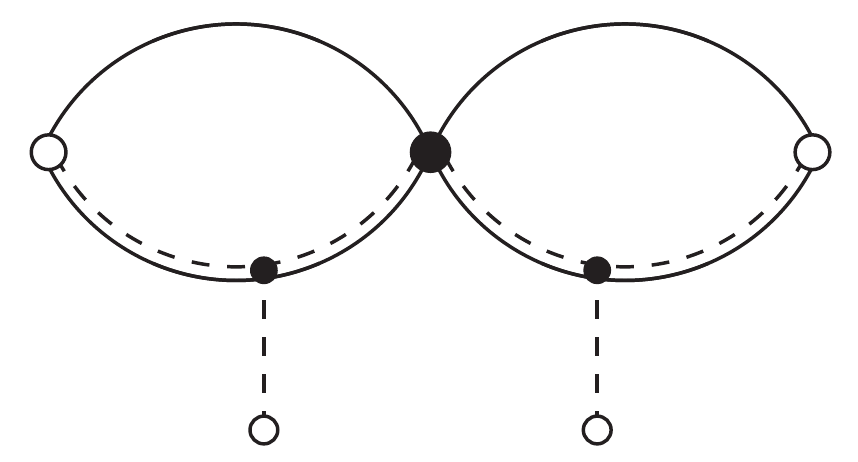}
\caption{
Two-loop diagram for the $D^\ast D$ self energy from pion forward scattering.
The $D^\ast$ flavors are constrained by the vertices to be $D^{\ast a}$, 
so the forward-scattered pion must be $\pi^0$.
}
\label{fig:D*Dselenergy2loop}
\end{figure}

The only two-loop diagram for the $D^\ast D$ self energy at NLO  is shown in Fig.~\ref{fig:D*Dselenergy2loop}.
The pion momentum can be routed through two $D^*$ propagators and through the contact vertex that attaches the two loops.
The Feynman rule for the $D^\ast D$ contact vertex in the $ab$ channel is $i\, C_0\, \delta^{mn}$,
where $m$ and $n$ are the vector indices of the attached $D^\ast$ lines.
The NLO contribution from this diagram is obtained by setting the small energies  to 0
in the two charm-meson propagators carrying  the pion momentum.
The integral over the loop momentum can be reduced to Eq.~\eqref{int1},  which is ultraviolet divergent.
The resulting contribution to the $D^\ast D$ self energy is
\beq
\Sigma_5^\mathrm{(NLO)}   =
- \frac{g_\pi^2}{3 f_\pi^2} \,  \mathfrak{n}_\pi 
\left\langle  \frac{q^2 }{\omega_{0q}^3} \right\rangle  
\frac{C_0}{2\pi/\mu} \big[ \Lambda - S_0(E_\mathrm{cm}) \big]^2.
\eeq
If we use Eq.~\eqref{C-Lambda} to replace $C_0/(2 \pi/\mu)$ by $1/(\Lambda - \gamma_X)$
and then take the large $\Lambda$ limit, the 2-loop contribution to the $D^\ast D$ self energy at NLO reduces to
\beq
\Sigma_5^\mathrm{(NLO)}  =
-\frac{g_\pi^2}{3 f_\pi^2} \,  \mathfrak{n}_\pi 
\left\langle  \frac{q^2 }{\omega_{0q}^3} \right\rangle  
\big[ \Lambda - 2 \, S_0(E_\mathrm{cm}) + \gamma_X\big].
\label{Sigma5NLO}
\eeq

\subsection{Contact vertex corrections}

The ultraviolet  divergent part of the $D^\ast D$ self energy at NLO
is the sum of the terms proportional to $\Lambda$ in Eqs.~\eqref{SigmaNLO} and Eq.~\eqref{Sigma5NLO}:
\bqa
\Sigma^\mathrm{(NLO)}_\mathrm{UV}&=&
\frac{g_\pi^2}{f_\pi^2} \,  \mathfrak{n}_\pi \, \Lambda 
\left\{ \frac{M}{M_X} \sum_c (2-\delta_{cb})\,   \mathcal{G}_{1,cb}
+  \frac{M_\ast}{3M_X} \sum_d (2-\delta_{ad}) \,  \mathcal{G}_{1,ad}^\ast  \right.
\nonumber\\
&& \hspace{2cm} \left.
+ \frac{1}{3}  \sum_c (2-\delta_{ac}) \left\langle  \frac{q^2}{\omega_{acq}^3} \right\rangle 
+  \frac{1}{3}(2-\delta_{ab})  \mathcal{H}_{ab} 
- \frac{1}{3} \left\langle  \frac{q^2 }{\omega_{0q}^3} \right\rangle \right\} .
\label{SigmaNLO-UV}
\eqa
These ultraviolet divergences must be canceled by ultraviolet divergences from corrections to the contact vertex.

The contact vertex $C_1$ required to compensate for $D$ and $D^\ast$ propagator corrections at short distances 
as well as short-distance vertex corrections is given in Eq.~\eqref{C1-C0}.
The expansion of $(2 \pi/\mu)/C_1$  to NLO is
\beq
\frac{2\pi/\mu}{C_1}  = \frac{2\pi/\mu}{C_0}
\left[ 1  + \delta Z_{\ast a} -  \frac{M}{M_X} \zeta_{\ast a} + \delta Z_b - \frac{M_\ast}{M_X}  \zeta_b 
- \frac{\delta C}{C_0} \right].
\label{C1-expand}
\eeq
The NLO corrections $\delta Z_b$ and $\delta Z_{\ast a}$  to the charm-meson residue factors 
are given in Eqs.~\eqref{deltaZDb} and \eqref{deltaZD*a}.
The NLO corrections to the charm-meson kinetic masses are given by
$\zeta_b$ and $\zeta_{\ast a}$ in Eqs.~\eqref{deltazetaDb} and \eqref{deltazetaD*a}.
The $\delta C$ term allows for additional corrections 
to the contact vertex that might be needed to compensate for short-distance vertex corrections.

\begin{figure}[t]
\includegraphics[width=0.8\textwidth]{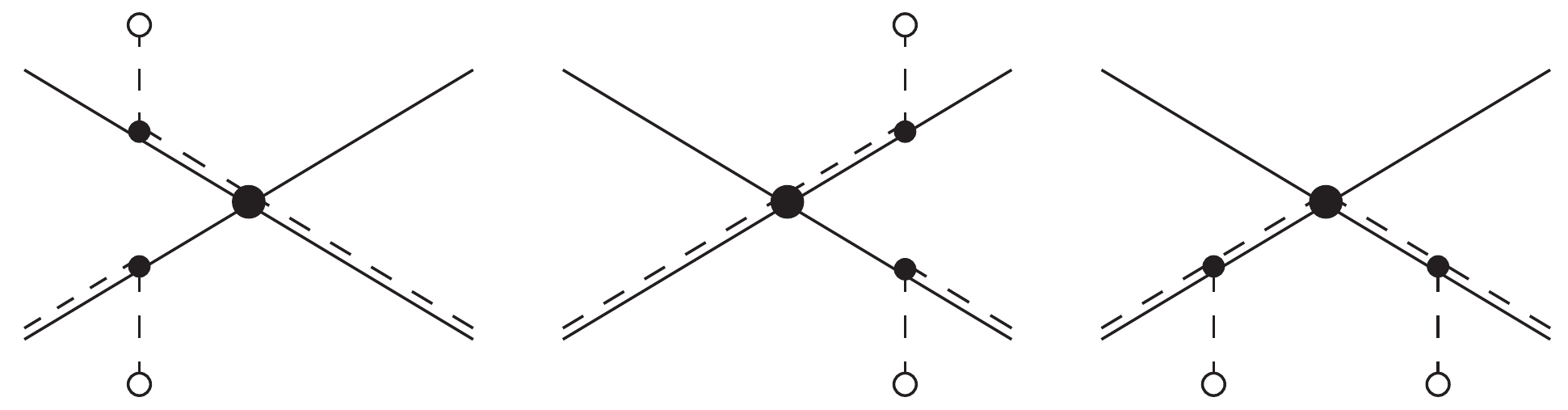}
\caption{
Diagrams for vertex corrections to the $D^\ast D$ contact interaction from pion forward scattering.
}
\label{fig:D*Dvertexcorrection}
\end{figure}

At  NLO in the heavy-meson expansion, there are $D^\ast D$ vertex corrections 
from pion forward scattering from the diagrams in Fig.~\ref{fig:D*Dvertexcorrection}.
The diagrams have two charm-meson propagators that carry a large energy of order $m_\pi$
that flows through the $D^\ast D$ contact vertex.
The NLO contribution from those diagrams is obtained 
by setting the small energies of order $m_\pi^2/M$ in the denominators of  those propagators equal to 0.
The changes in the $D^\ast D$ contact vertex required to cancel the vertex corrections 
from the first two diagrams  in Fig.~\ref{fig:D*Dvertexcorrection} are
\bqa
\delta C_1 = \delta C_2  =
 \frac{g_\pi^2}{6f_\pi^2} \,  C_0\, \mathfrak{n}_\pi \, (2-\delta_{ab})\, \mathcal{H}_{ab} .
\eqa
The change in the $D^\ast D$ contact vertex required to cancel the vertex correction from the third diagram  in Fig.~\ref{fig:D*Dvertexcorrection} is
\beq
\delta C_3  =
- \frac{g_\pi^2}{3f_\pi^2} \,  C_0 \,  \mathfrak{n}_\pi 
\left\langle  \frac{q^2 }{\omega_{0q}^3} \right\rangle  .
\eeq
Upon using Eq.~\eqref{C-Lambda} to replace the factor of $(2\pi/\mu)/C_0$ 
in Eq.~\eqref{C1-expand}  by $\Lambda - \gamma_X$,
we obtain terms linear in the ultraviolet cutoff $\Lambda$:
\bqa
\frac{2\pi/\mu}{C_1}  &=& \big[ \Lambda - \gamma_X \big]
\left\{ 1 -  \frac{g_\pi^2}{f_\pi^2} \,  \mathfrak{n}_\pi
\left[ \frac{M}{M_X} \sum_c (2-\delta_{cb})\,   \mathcal{G}_{1,cb} 
+  \frac{M_\ast}{3 M_X} \sum_d (2-\delta_{ad})   \mathcal{G}_{1,ad}^\ast \right. \right.
\nonumber\\
&& \hspace{3cm}  \left. \left.
+  \frac13 \sum_c (2-\delta_{ac}) \left\langle  \frac{q^2}{\omega_{acq}^3} \right\rangle
+  \frac13 (2-\delta_{ab})  \mathcal{H}_{ab} 
-  \frac13 \left\langle  \frac{q^2}{\omega_{0q}^3} \right\rangle \right] \right\}.
\label{C1coeffs}
\eqa
The terms proportional to $\Lambda$ cancel the ultraviolet divergences in Eqs.~\eqref{SigmaNLO-UV}.

\subsection{$\bm{D^\ast D}$ Inverse Propagator}
\label{sec:D*DinvProp}

The denominator of the complete $D^\ast D$ propagator  has the form
\beq
(2 \pi/\mu)/C_1 -\Lambda + S_0(E_\mathrm{cm}) + \Sigma(E_\mathrm{cm},P),
\label{D*DinvProp}
\eeq
where $C_1$ is the complete contact-interaction vertex 
and $\Sigma(E_\mathrm{cm},P)$ is the $D^\ast D$ self energy.
The $D^\ast D$ self energy at NLO is the sum of 
Eqs.~\eqref{SigmaXLO}, \eqref{SigmaNLO}, and \eqref{Sigma5NLO}.
The ultraviolet divergences in the $D^\ast D$ self energy are canceled by the corrections 
 from the contact interaction vertex $C_1$ in Eq.~\eqref{C1coeffs}.
The net effect is essentially to replace $\Lambda$ in the $D^\ast D$ self energy by $\gamma_X$.
The terms in the $D^\ast D$ self energy proportional to $1/S_0(E_\mathrm{cm})$, 
which diverge at the branch point of $S_0(E_\mathrm{cm})$,
can be absorbed by replacing $S_0(E_\mathrm{cm})$ in Eq.~\eqref{D*DinvProp} by 
the square-root function $S_1(E_\mathrm{cm},P)$ in Eq.~\eqref{S1-E}.
An appropriate resummation of higher order corrections would replace the remaining  $S_0(E_\mathrm{cm})$ terms
 in the $D^\ast D$ self energy by $S_1(E_\mathrm{cm},P)$.
 The resulting expression for the denominator of the $D^\ast D$ propagator at NLO is
\bqa
&\big[ -\gamma_X + S_1(E_\mathrm{cm},P) \big]&
\Bigg\{  1 -\frac{g_\pi^2}{f_\pi^2} \,  \mathfrak{n}_\pi 
\left[ \frac{M}{M_X} \sum_c (2-\delta_{cb})  \mathcal{G}_{1,cb}
+ \frac{M_\ast}{3 M_X} \, 
\sum_d(2-\delta_{ad}) \mathcal{G}_{1,ad}^* \right.
\nonumber\\
&&\hspace{1cm}
\left. +\frac{1}{3} \, 
  \sum_c (2-\delta_{ac}) \left\langle  \frac{q^2}{\omega_{acq}^3} \right\rangle  
+ \frac{1}{3} \,  (2-\delta_{ab})\, \mathcal{H}_{ab}
- \frac{2}{3} \left\langle  \frac{q^2}{\omega_{0q}^3} \right\rangle  \right] \Bigg\} 
\nonumber\\
&&\hspace{-3cm}
+ \frac{g_\pi^2 \Delta}{4f_\pi^2 M_X} \,  \mathfrak{n}_\pi \, S_1(E_\mathrm{cm},P) 
\left[ \sum_c (2-\delta_{cb}) \mathcal{G}_{1,cb}
-\frac13 \sum_d (2-\delta_{ad}) \mathcal{G}_{1,ad}^*  \right].
\label{D*DPropNLO}
\eqa


\section{Thermal energy of the molecule}
\label{sec:Compare}

In this Section, we use the  $D^\ast D$ self-energy at NLO in the heavy-meson expansion
to determine the thermal correction to the binding momentum 
of a loosely bound charm-meson molecule in the pion gas.
We then calculate the thermal mass shifts and thermal widths of $X(3872)$ and $T_{cc}^+(3875)$
in the pion gas.

\subsection{Correction to the binding momentum}
\label{sec:deltagamma}

The denominator of the $D^\ast D$ propagator through NLO in Eq.~\eqref{D*DPropNLO}
 can be expressed in the form of the denominator in Eq.~\eqref{Xprop-branch}
by factoring out the coefficient of $S_1$ into a multiplicative factor $Z_X^{-1}$ given by
\bqa
Z_X^{-1} &=& 1- \frac{g_\pi^2}{f_\pi^2} \,  \mathfrak{n}_\pi 
\left[   \frac{M}{M_X} \left(1 - \frac{\Delta}{4 M} \right) \sum_c (2-\delta_{cb})  \mathcal{G}_{1,cb}
+  \frac{M_\ast}{3 M_X} \left(1+\frac{\Delta}{4M_\ast} \right)
\sum_d(2-\delta_{ad}) \mathcal{G}_{1,ad}^* \right.
\nonumber\\
&&\hspace{2cm}
\left. +\frac{1}{3} \, 
  \sum_c (2-\delta_{ac}) \left\langle  \frac{q^2}{\omega_{acq}^3} \right\rangle  
+ \frac{1}{3} \,  (2-\delta_{ab})\, \mathcal{H}_{ab}
  -\frac23 \left\langle  \frac{q^2}{\omega_{0q}^3} \right\rangle  \right] .
\label{ZX}
\eqa
After factoring out $Z_X^{-1}$ and re-expanding to NLO, 
the denominator of the $D^\ast D$ propagator  reduces to the much simpler expression
\beq
 -\gamma_X  \left[ 1 - \frac{g_\pi^2 \Delta}{4f_\pi^2 M_X} \,  \mathfrak{n}_\pi
\left( \sum_c (2-\delta_{cb}) \mathcal{G}_{1,cb}
-\frac13 \sum_d (2-\delta_{ad}) \mathcal{G}_{1,ad}^*  \right) \right] + S_1(E_\mathrm{cm},P).
\label{D*DPropNLO2}
\eeq
This has the same form as the denominator of the amplitude for a loosely bound molecule in 
Eq.~\eqref{sumbubbles} but with $S_0(E_\mathrm{cm})$ 
replaced by the thermally modified square-root function $S_1(E_\mathrm{cm},P)$ in Eq.~\eqref{S1-E}
and with a correction $\delta \gamma_X$ to the binding momentum.
The NLO thermal correction to the binding momentum is
\beq
\delta \gamma_X = 
- \frac{g_\pi^2 \Delta}{4f_\pi^2 M_X} \,  \mathfrak{n}_\pi \, \gamma_X 
\left( \sum_c (2-\delta_{cb}) \mathcal{G}_{1,cb}
-\frac13 \sum_d (2-\delta_{ad}) \mathcal{G}_{1,ad}^*  \right).
\label{gammaXNLO}
\eeq

The residue factor  $Z_X^{-1}$ in Eq.~\eqref{ZX} can be absorbed into the normalization 
of the local composite operator whose 2-point Green function is the $D^\ast D$ propagator.  
The NLO corrections in $Z_X^{-1}$ therefore have no physical effects.
The remaining NLO corrections to the $D^\ast D$ self energy
have been completely absorbed into thermal corrections to parameters of ZREFT.
The specific parameters are those that determine the propagators of the individual charm mesons at NLO,
which appear in the square-root function $S_1(E_\mathrm{cm},P)$ in Eq.~\eqref{S1-E},
and the complex binding momentum, whose thermal correction is given in Eq.~\eqref{gammaXNLO}.

The fractional correction  $\delta \gamma_X/\gamma_X$ to the binding momentum from Eq.~\eqref{gammaXNLO}
is different for $X(3872)$ with constituents $D^{*0} \bar D^0$/$D^0 \bar D^{*0}$
and for $T_{cc}^+(3875)$  with constituents $D^{*+} D^0$.
For $X$, the combination of thermal averages in $\delta \gamma_X$ is
$G_{1,00}  + 2 G_{1,+0} -  (G_{1,00}  + 2 G_{1,0+})^*/3$.  
For a pion gas at $T_\mathrm{kf} = 115$\,MeV,
 the fractional correction to the binding momentum is
 $\delta \gamma_X/\gamma_X = (-3.0+2.6\, i)\times 10^{-4}$.  
For $T_{cc}^+$, the combination of thermal averages in Eq.~\eqref{gammaXNLO}
is $G_{1,00}  + 2 G_{1,+0} -  (G_{1,++}  + 2 G_{1,+0})^*/3$. 
For a pion gas at $T_\mathrm{kf} = 115$\,MeV, the fractional correction to the binding momentum is 
$\delta \gamma_T/\gamma_T=  (-2.9 + 3.1 \, i)\times 10^{-4}$. 
 
The thermal averages in Eq.~\eqref{gammaXNLO} can be expressed as double expansions 
in isospin splittings divided by $m_\pi$ and in pion/charm-meson mass ratios.
The leading term in the  real part of the fractional change in the binding momentum is
\beq
\mathrm{Re} \big[\delta \gamma_X/\gamma_X\big] \approx
- \frac{g_\pi^2 \Delta}{2f_\pi^2 M_X} \,  \mathfrak{n}_\pi 
\left\langle  \frac{\omega_q^2 + m_\pi^2}{q^2 \omega_q} \right\rangle  .
\label{Re[gammaXNLO]}
\eeq
[For a pion gas at $T_\mathrm{kf} = 115$\,MeV, its value is
$\mathrm{Re}[\delta \gamma_X/ \gamma_X] = -3.4 \times 10^{-4}$. 
The more accurate results from Eq.~\eqref{gammaXNLO} differ by
$-11\%$  for $X$ and by $-15$\%  for $T_{cc}^+$.
The leading term in the imaginary part of the fractional change in the binding momentum is
\bqa
\mathrm{Im} \big[\delta \gamma_X/\gamma_X\big] &\approx& 
 \frac{3g_\pi^2 m_\pi \Delta}{16\pi f_\pi^2 M_X} \,   \mathfrak{f}_\pi(m_\pi)
\left( \sum_c (2-\delta_{cb}) \, q_{cb} \,  \theta \big(  \Delta_{cb} - m_{\pi cb} \big) 
\right.
\nonumber\\
&&\hspace{4cm} \left.
+\frac13 \sum_d (2-\delta_{ad}) \, q_{ad}\,  \theta \big(  \Delta_{ad} - m_{\pi ad} \big)   \right).
\label{Im[gammaXNLO]}
\eqa
where $q_{cd} = \sqrt{\Delta_{cd}^2 - m_{\pi cd}^2}$ is the momentum of the pion in the decay
$D^{\ast c} \to D^d \pi$.
This expression is different for $X(3872)$ with constituents $D^{*0} \bar D^0$/$D^0 \bar D^{*0}$
and for $T_{cc}^+(3875)$  with constituents $D^{*+} D^0$.
For $X$, the linear combination of pion momenta in Eq.~\eqref{Im[gammaXNLO]} is 
$(4q_{00}  + 6 q_{+0})/3$. 
For a pion gas at  $T_\mathrm{kf} = 115$\,MeV,
the imaginary part of the fractional change in the binding momentum is
$\mathrm{Im}[\delta \gamma_X/ \gamma_X ]= 2.8 \times 10^{-4}$. 
For $T_{cc}^+$, the linear combination of pion momenta  in Eq.~\eqref{Im[gammaXNLO]} is
$(3q_{00} + q_{++}  + 8 q_{+0})/3$. 
For a pion gas at  $T_\mathrm{kf} = 115$\,MeV, 
the imaginary part of the fractional change in the binding momentum is
$\mathrm{Im}[\delta \gamma_T/ \gamma_T] = 3.4 \times 10^{-4}$. 
The more accurate results for $\mathrm{Im}[\delta \gamma_X/\gamma_X]$ from Eq.~\eqref{gammaXNLO} 
differ from Eq.~\eqref{Im[gammaXNLO]} by about $-8\%$  for both $X$ and  $T_{cc}^+$. 

The only thermal average in the inverse propagator in Eq.~\eqref{D*DPropNLO}
that does not have a well defined limit as $\epsilon \to 0^+$ is $\mathcal{H}_{ab}$,
which has a term that diverges as $1/\epsilon$  in this limit.
A more careful treatment of this thermal average would regularize the
divergence by replacing $\epsilon$ by appropriate decay widths of charm mesons, 
as in Eq.~\eqref{Hab-Gamma} of Appendix~\ref{app:PionIntegral}.
The thermal average $\mathcal{H}_{ab}$ is therefore sensitive to these decay widths.
Since $\mathcal{H}_{ab}$ appears in the residue factor $Z_X^{-1}$ in Eq.~\eqref{ZX} 
but not in the denominator of the $D^\ast D$ propagator in Eq.~\eqref{D*DPropNLO2},
it does not affect the energy-momentum relation for $X$.
We therefore do not give any analytic approximation for  $\mathcal{H}_{ab}$.

\subsection{Thermal mass shift and thermal width}
\label{sec:deltavarepsilonX}

The zero of the inverse propagator in Eq.~\eqref{D*DPropNLO2}
determines the energy-momentum relation for $X$.
The pole energy of $X$ with zero 3-momentum at NLO can be expressed as
\beq
E_X = ( \varepsilon_{\ast a} - i\,  \Gamma_{\ast a}/2) + \varepsilon_b  + ( \delta\varepsilon_{\ast a} +  \delta\varepsilon_b)
 - (\gamma_X + \delta \gamma_X)^2/(2\mu) ,
\label{EX-gammaXNLO} 
\eeq
where $\delta \gamma_X$ is the NLO correction to the binding momentum in Eq.~\eqref{gammaXNLO}.
We have taken into account the decay width of $D^{\ast a}$ at $T=0$
by replacing $i\epsilon$ in Eq.~\eqref{S1-E} with $i\Gamma_{\ast a}/2$.

For $X(3872)$, the pole energy at $T=0$ relative to the real $D^{\ast 0} \bar D^0$ threshold 
determined in Ref.~\cite{LHCb:2020xds} is $(0.025 - 0.140\, i)$\,MeV. 
In a pion gas at  $T_\mathrm{kf} = 115$\,MeV, the pole energy 
relative to the real $D^{\ast 0} \bar D^0$ threshold from Eq.~\eqref{EX-gammaXNLO}  is
$(+1.64 - 0.21 \,i)$\,MeV. 
The  change in the pole energy comes primarily from the complex thermal energy shifts of the  
$D^{*0} \bar D^0$/$D^0 \bar D^{*0}$ constituents, whose sum  is $(+1.61 - 0.07\,i)$\,MeV. 
The contribution from the complex thermal correction $\delta \gamma_X$ to the binding momentum is 
$(+0.04 + 0.08\,i)$\,keV. 
Its real and imaginary parts are smaller than those from the energy shifts of the constituents by orders of magnitude. 

For $T_{cc}^+(3875)$, the pole energy at $T=0$ relative to the real $D^{\ast +} D^0$ threshold 
determined in Ref.~\cite{LHCb:2021auc} is $(-0.36 -0.024\, i)$\,MeV. 
In a pion gas at  $T_\mathrm{kf} = 115$\,MeV, the pole energy 
relative to the real $D^{\ast +} D^0$ threshold from Eq.~\eqref{EX-gammaXNLO} is 
$(+1.20 - 0.10\,i)$\,MeV. 
The change  in the pole energy comes primarily from the complex thermal energy shifts of the  
$D^{*+} D^0$ constituents, whose sum is $(+1.56 - 0.08 \,i)$\,MeV. 
The contribution from the complex thermal  correction $\delta \gamma_T$ to the binding momentum  is
$(+0.20 - 0.23\,i)$\,keV. 
Its real and imaginary parts are smaller than those from the energy shifts of the constituents
by orders of magnitude.  

\begin{figure}[t]
\includegraphics[width=0.48\textwidth]{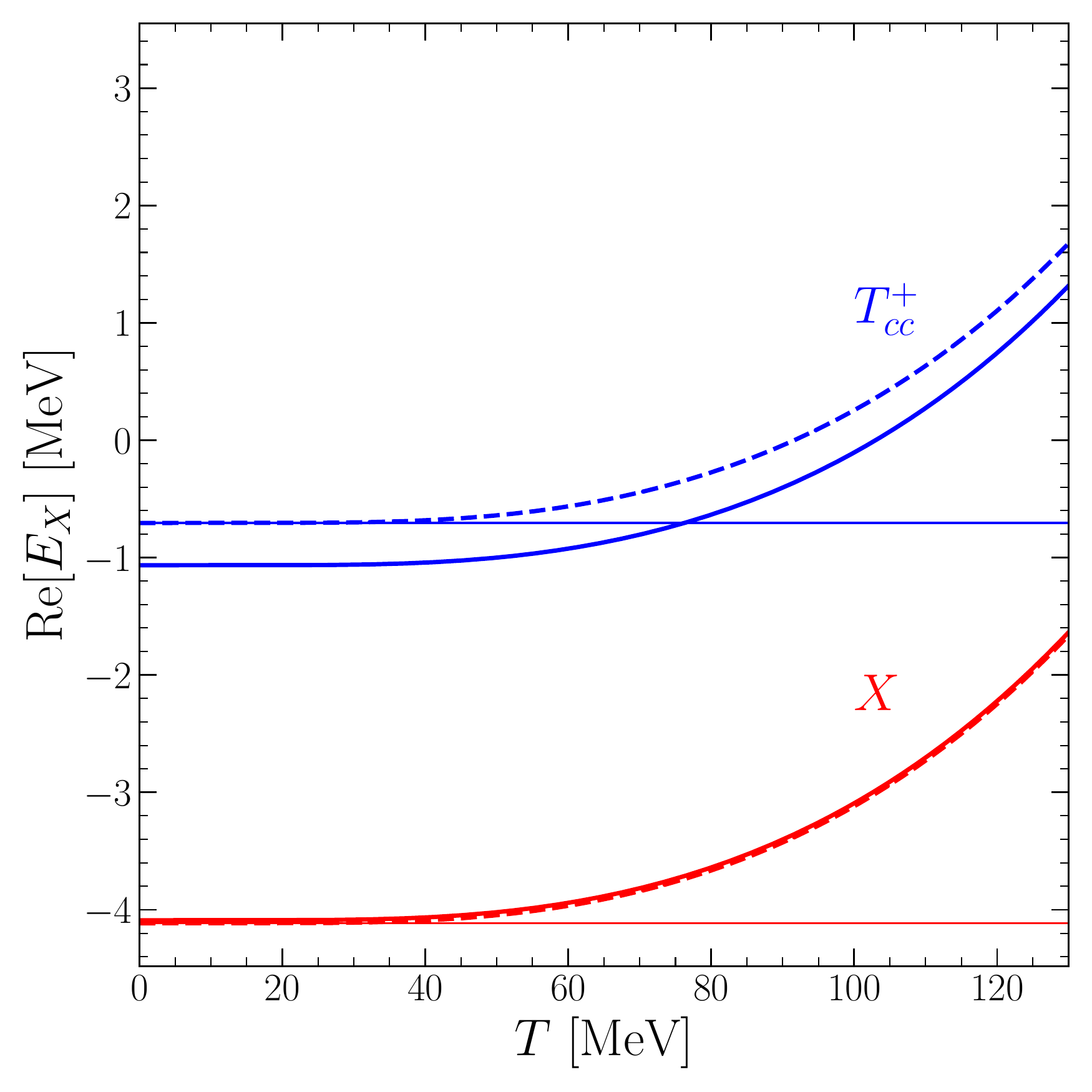} ~
\includegraphics[width=0.48\textwidth]{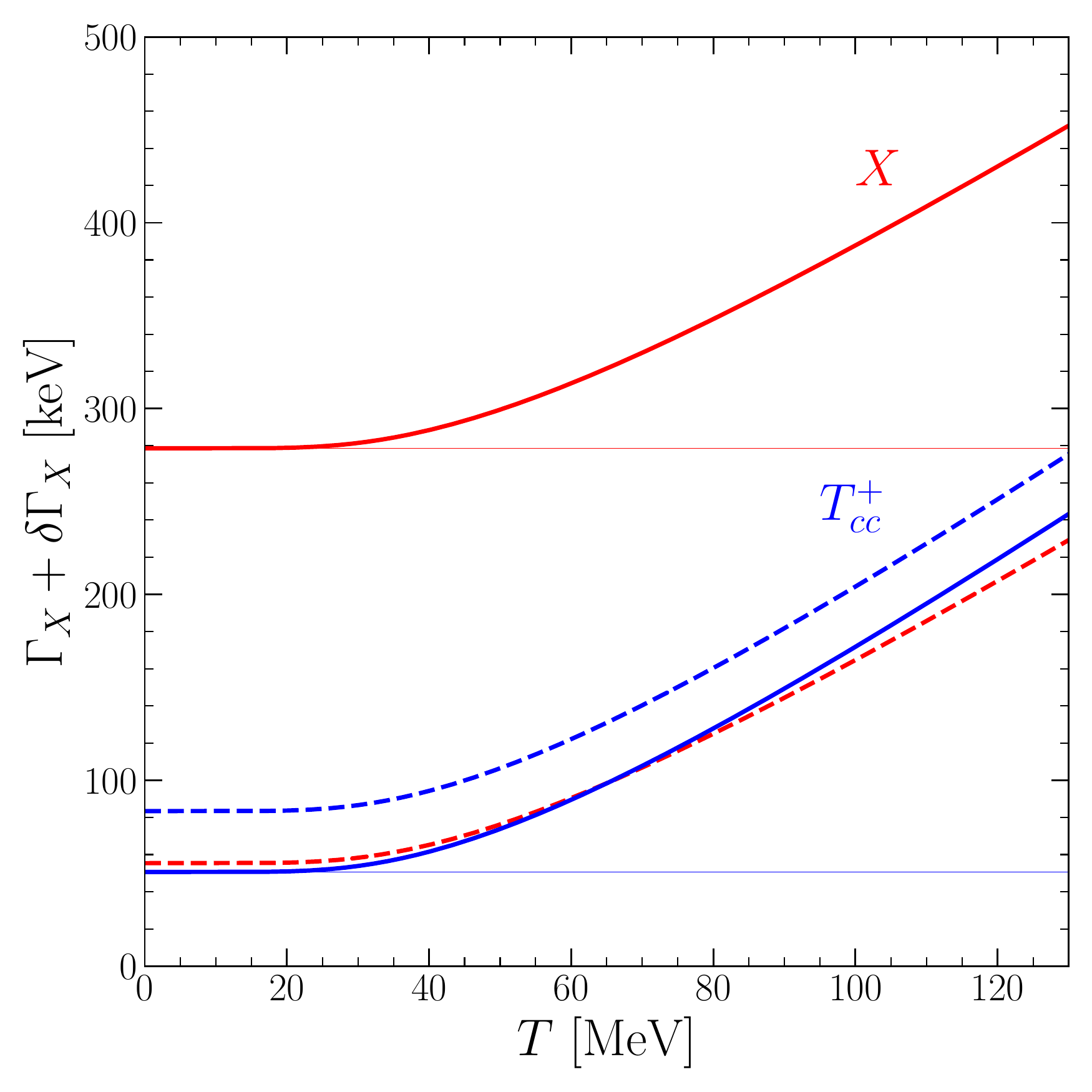}
\caption{
The real parts of the poles of the pair propagators (left panel) and the thermal widths (right panel) for $X(3872)$, $T_{cc}^+(3875)$, and their constituents in the pion gas as functions of the temperature $T$.
In the left panel, the solid curves are the sums of the rest energy at $T=0$ and the thermal mass shift
for $X$ (red lower curve) and $T_{cc}^+$ (blue upper curve).
The dashed curves and the horizontal lines are the  corresponding charm-meson pair thresholds 
in the pion gas at temperature $T$ and at $T=0$, respectively.
In the right panel, the solid curves are the sums of the decay width at $T=0$ and the thermal width
for $X$ (red upper curve) and $T_{cc}^+$ (blue lower curve).
The horizontal lines are the decay widths
of $X$ (red upper line) and $T_{cc}^+$ (blue lower line)  at $T=0$ MeV. 
The dashed curves are the sums of the widths of the constituents $D^{\ast 0}$ and $D^0$ of $X$ (red lower curve) 
and  the constituents $D^{\ast +}$ and $D^0$ of $T_{cc}^+$ (blue upper curve).
}
\label{fig:DD*propagatorenergyshift}
\end{figure}

The  temperature dependence of the pole energies of $X(3872)$ and $T_{cc}^+(3875)$
is illustrated in Fig.~\ref{fig:DD*propagatorenergyshift}.
The range of the temperature $T$ extends  only up to 130\,MeV,
 which may already be beyond the range of validity of our results. 
In the left panel of Fig.~\ref{fig:DD*propagatorenergyshift}, a solid curve 
is the real part of $E_X$, which is the sum of the rest energy of the molecule at $T=0$ and its thermal mass shift.
A dashed curve  is the $T$-dependent charm-meson pair threshold.
The thermal mass shifts of $X$ and $T_{cc}^+$ increase monotonically  with $T$.
Their binding energies relative to the charm-meson pair thresholds in the pion gas  
are the differences between the dashed and solid curves.
The $T$-dependence of the binding energies, which comes from the thermal correction to the binding momentum,
 is too small to be visible in Fig.~\ref{fig:DD*propagatorenergyshift}.
In the right panel of Fig.~\ref{fig:DD*propagatorenergyshift}, 
a solid curve is the imaginary part of $E_X$ multiplied by $-2$, 
which is the sum of the decay width of the molecule  and its thermal width.
The thermal widths increase monotonically with $T$. 
Their dependence on $T$ comes almost entirely from the thermal rest energies of the charm-meson constituents.
 The remaining $T$-dependence from the complex thermal correction to the binding momentum
is too small to be visible in  Fig.~\ref{fig:DD*propagatorenergyshift}.

\subsection{Comparison with Previous Work}
\label{sec:Xcompare}

There have been two previous calculations of the  thermal energies of $X(3872)$ in a pion gas \cite{Cleven:2019cre,Montana:2022inz}. 
The  temperatures ranged from 0 to 150\,MeV.
Our results are not applicable at temperatures as high as 150\,MeV, which is near the hadronization temperature.
We will therefore only  compare our results with the previous calculations at 100\,MeV.
The comparisons are presented in Table~\ref{tab:X-comparison}.

The effect of a  hot pion bath on the $X(3872)$
has been studied previously by Cleven, Magas, and Ramos (CMR) \cite{Cleven:2019cre}.
The $X$ was  identified with a dynamically generated bound state with energy 2.5\,MeV below the $D^\ast D$ threshold.
This energy is the pole in a unitarized $DD^\ast$ amplitude given by the analytic solution of  a 
regularized Lippmann-Schwinger equation.  The $DD^\ast$ tree amplitude in that equation is the vertex 
from an effective Lagrangian for pseudoscalar and vector mesons with an approximate $SU(4)$ symmetry,
as in Ref.~\cite{Cleven:2017fun}.
The thermal $D^\ast D$ amplitude $T_{DD^\ast}$ is obtained by the numerical solution of the 
Lippmann-Schwinger integral equation with charm meson propagators 
that include the self-consistent thermal $D^{(\ast)}$ self energies calculated in Ref.~\cite{Cleven:2017fun}.
The energy of $X$ and its width were identified with 
the energy at the peak of $|T_{DD^\ast}|$ and the width of the peak, respectively.
As $T$ increases from 0 to 100\, MeV, 
the energy of $X$ increases from $-2.5$ to +3\, MeV. 
Although it was not stated in Ref.~\cite{Cleven:2019cre}, the $D^\ast D$ threshold was actually
held constant at its $T=0$ value, according to Ref.~\cite{Montana:2022inz}. 
The  thermal energy of $X$ in Ref.~\cite{Cleven:2019cre}
should therefore be interpreted as the difference between  its
thermal energy and an unknown $T$-dependent shift in the  $D^\ast D$ threshold.
As $T$ increases from 0 to 100\,MeV, 
the width of $X$ increases from 0 to 30\,MeV.

The effect of a  hot pion bath on the $X(3872)$ has also been studied previously 
by Monta\~na, Ramos, Tolos, and Torres-Rincon (MRTT) \cite{Montana:2022inz}.
They also considered its effect on a possible P-wave charm-meson molecule called $X(4140)$.
The $X$ was identified with a dynamically generated bound state with energy 4.1\,MeV below the $D^\ast D$ threshold.
This is the energy of a pole in a unitarized $DD^\ast$ amplitude given by the analytic solution of  a 
regularized Lippmann-Schwinger equation.  The $DD^\ast$ tree amplitude in that equation is obtained
from the effective Lagrangian for pseudoscalar and vector mesons with an approximate $SU(4)$ symmetry
used in Ref.~\cite{Montana:2020vjg}.
The thermal $D^\ast D$ amplitude $T_{DD^\ast}$ is obtained  from the numerical solution of the 
Lippmann-Schwinger integral equation with charm meson propagators 
that include the self-consistent thermal $D^{(\ast)}$ self energies calculated in Ref.~\cite{Montana:2020vjg}.
The energy of $X$ and its width were identified with the energy of the peak 
in  $-\mathrm{Im}[T_{DD^\ast}]$ and the width of the peak, respectively.
As $T$ increases from 0 to 100\,MeV,
the energy of $X$ decreases from $-4$ to $-30$\,MeV.
Most of the change comes from the thermal mass shifts of the charm-meson constituents,
whose sum decreases from 0 to $-27$\,MeV. 
The energy of $X$ relative to the $T$-dependent charm-meson-pair threshold
increases from $-4$ to $-3$\,MeV. 
As $T$ increases from 0 to 100\,MeV, the width of $X$ increases 
from 0 to 30\,MeV.
Most of the change comes from the thermal widths of the charm-meson constituents,
whose sum increases from 0 to 34\,MeV. 
The thermal width agrees with that in Ref.~\cite{Cleven:2019cre} at 100\,MeV. 

\begin{table}
    \centering
    \begin{tabular}{ccccccc}
        & \multicolumn{2}{c}{$T = 0$} & \multicolumn{4}{c}{$T = 100$~MeV} \\
        \cmidrule(lr){2-3} \cmidrule(lr){4-7}
        & Re$[E_X]$ & $\Gamma_X$ & Re$[E_X]$ & Re$[E_X - \delta\varepsilon_0 - \delta\varepsilon_{\ast0}]$ & 
        $\delta\Gamma_X$ & $\delta \Gamma_X - \delta \Gamma_0 - \delta\Gamma_{\ast0}$ \\
         \toprule
        Cleven {\it et al.} \cite{Cleven:2019cre} & $-2.5$ & $0.0$  & &  $+3.0$ & 30. & +5.\\
        \cmidrule(lr){2-3} \cmidrule(lr){4-7}
        Monta\~na {\it et al.} \cite{Montana:2022inz} & $-4.1$ & $0.0$  & $-30.$ & $-3.0$ & $30.$ & $-4.$ \\
      \hline
        This work & $+0.025$ & 0.28  & +1.02 & $+0.025$ & 0.109 & $-0.0001$\\
       \bottomrule
    \end{tabular}
    \caption{
    The real part of the pole of the pair propagator $\text{Re}[E_X(T)]$, the real part relative to the $T$-dependent $D^{*0}D^{0}$ threshold $\delta\varepsilon_0(T) + \delta\varepsilon_{\ast0}(T)$, the thermal width $\delta \Gamma_X(T)$, and the thermal width relative to the sum $\delta \Gamma_{0}(T) + \delta \Gamma_{\ast0}(T)$ of the thermal widths of the constituents for $X(3872)$ at $T=0$ and $T=100$ MeV.
    The energies and widths are in MeV.}
    \label{tab:X-comparison}
\end{table}

We proceed to compare our results for $X(3872)$ with those in MRTT for the pion gas at $T=100$\,MeV.
Our result in Table~\ref{tab:X-comparison} for
the sum of the thermal mass shifts of the constituents $D^{*0}$ and $\bar D^0$
from Eqs.~\eqref{deltaepsilonDb} and \eqref{deltaepsilonD*a}  is +1.00\,MeV.
This is an order of magnitude smaller than the sum of the thermal mass shifts of $D^\ast$ and $\bar D$ 
in Ref.~\cite{Montana:2020vjg} and it has the opposite sign.
Our result in Table~\ref{tab:X-comparison} for
the real part of the pole energy of $X$ relative to the complex charm-meson-pair threshold from Eq.~\eqref{EX-gammaXNLO} is
$+25$\,keV  at $T=0$ and it increases by only 0.03\,keV at $T=100$\,MeV.
 The corresponding energy  at $T=100$\,MeV in MRTT
is about two orders of magnitude larger and it has the opposite sign.
Our result in Table~\ref{tab:X-comparison} for
the sum of the  thermal widths of the constituents $D^{*0}$ and $\bar D^0$
from Eqs.~\eqref{deltaepsilonDb} and \eqref{deltaepsilonD*a} is 0.11\,MeV.
This is smaller than the sum of the thermal widths of $D^\ast$ and $\bar D$ in Ref.~\cite{Montana:2020vjg}
by a factor of about 300. 
Our result in Table~\ref{tab:X-comparison} for
the difference between the width of $X$ from Eq.~\eqref{EX-gammaXNLO} 
and the sum of the widths of the constituents $D^{*0}$ and $\bar D^0$ is 
$225$\,keV at $T=0$ and it decreases by only 0.11\,keV  at $T=100$\,MeV.
The corresponding difference in MRTT is 0 at $T=0$ and it
decreases to $-4$\,MeV  at $T=100$\,MeV.

The orders-of-magnitude discrepancies between our results for $X$ and its constituents
and those in Ref.~\cite{Montana:2022inz} are a serious issue.
We cannot exclude the possibility  that higher orders in the chiral expansion are much larger than the leading order, even at temperatures as low as 100\,MeV.
Another possibility is that  the proposed generalizations to nonzero  temperature of the integral equations in Ref.~\cite{Montana:2022inz} 
may give thermal mass shifts and thermal widths for the charm-meson molecule that are much too large at low temperatures.


\section{Summary}
\label{sec:Summary}

We have calculated the thermal energy of a loosely bound charm-meson molecule in a pion gas.
The molecule consists of a bound pair of charm mesons in a channel we denoted by $D^\ast D$.
It is associated with a pole in the $D^\ast D$ propagator very close to the $D^\ast D$ threshold.
At zero  temperature, the molecule can be described by a ZREFT for nonrelativistic charm mesons
with a nonperturbative contact interaction in the $D^\ast D$ channel.
At leading order in the ZREFT, the $D^\ast D$ propagator has the simple form in Eq.~\eqref{sumbubbles}.
We assumed that 
the pole in the $D^\ast D$ propagator and the branch cuts associated with the $D^\ast D$ threshold 
remain the only nearby  singularities even in the thermal environment provided by the pion gas.
The temperature of the pion gas was assumed to be low enough 
that the interactions between charm mesons and pions can be described by 
heavy-hadron $\chi$EFT at leading order.

The $D^\ast D$ self-energy was calculated to NLO in the heavy-meson expansion.
At LO in that expansion, the $D^\ast D$ self-energy is the sum of the contributions in Eqs.~\eqref{SigmaXLO} 
from two one-loop diagrams.
Both terms diverge at the  $D^\ast D$ threshold, but these diverging terms can be absorbed into the parameters of ZREFT
by adding the LO  thermal corrections $\delta \varepsilon$ and $\delta \varepsilon_\ast$ to the charm-meson rest energies 
$\varepsilon$ and $\varepsilon_\ast$ in the square-root function $S_0(E_\mathrm{cm})$ in Eq.~\eqref{S0-Ecm}.
Thus there are no thermal corrections to the binding energy of the mloecule at this order.

At NLO in the heavy-meson expansion, the $D^\ast D$ self-energy is much more complicated.
It is the sum of  the contributions in Eqs.~\eqref{SigmaNLO} from one-loop diagrams
and  in Eq.~\eqref{Sigma5NLO} from a two-loop diagram.
The $D^\ast D$ self-energy at NLO has linear ultraviolet divergences 
as well as terms that diverge at the  $D^\ast D$ threshold.
The  terms that diverge at the  $D^\ast D$ threshold can be absorbed into the parameters of ZREFT
by replacing $S_0(E_\mathrm{cm})$ in the $D^\ast D$ propagator
by the thermally corrected square-root function $S_1(E_\mathrm{cm},P)$ in Eq.~\eqref{S1-E},
which takes into account thermal corrections to the charm-meson rest energies
and their kinetic masses.
The ultraviolet divergences in the $D^\ast D$ self-energy at NLO are canceled by the
correction to the strength of the contact  vertex in Eq.~\eqref{C1coeffs}, which
compensates for thermal corrections to the charm-meson propagators as well as thermal vertex corrections. 

After canceling the ultraviolet divergences and absorbing the  divergences at the  $D^\ast D$ threshold
into charm-meson propagator corrections,
the denominator of the complete $D^\ast D$ propagator at NLO reduces to Eq.~\eqref{D*DPropNLO}.
This form is consistent with our assumption that the molecule in the pion gas can be described by ZREFT.
After factoring out the residue factor $Z_X^{-1}$ in Eq.~\eqref{ZX},
the denominator of the complete $D^\ast D$ propagator at NLO reduces to Eq.~\eqref{D*DPropNLO2}.
Aside from the thermal corrections to the charm-meson propagators that appear in the square-root function,
the only other thermal correction is the NLO correction to the binding momentum  in Eq.~\eqref{gammaXNLO}.
It is proportional to the pion number density $ \mathfrak{n}_\pi$, 
and it is suppressed by a factor of $\Delta/M_X \approx m_\pi/M_X$.
The thermal mass shift and thermal width of the molecule is therefore 
dominated by the thermal mass shifts  and thermal widths of the charm-meson constituents.

The thermal energy of the $X(3872)$ in a pion gas has been calculated previously 
in Ref.~\cite{Cleven:2019cre} (CMR) and Ref.~\cite{Montana:2022inz} (MRTT).
In MRTT, the thermal  mass shift and the thermal width of $X(3872)$ 
were calculated as a function of $T$ up to 150\,MeV.
Our results are certainly not valid at such a high  temperature,
but we can compare results  near the  kinetic freeze-out temperature. 
At  such a temperature, 
the thermal mass shift for $X$  comes almost entirely from the thermal mass shifts of the $D^* D$ constituents.
The sum of our thermal mass shifts for $D^*$ and $D$ is  positive
while the sum in MRTT is negative. 
At  $T=100$\,MeV, the sum in MRTT is more than an order of magnitude larger in absolute value than ours.
At  a temperature near kinetic freezeout, the thermal contribution to the width of $X$ 
comes almost entirely from the thermal widths of the $D^* D$ constituents.
At  $T=100$\,MeV,  the sum of the thermal widths for $D^*$ and $D$ in MRTT
is more than two orders of magnitude larger than ours.

Since the results for $X$ in CMR and MRTT are completely numerical, 
it is difficult to identify the reasons for the orders-of-magnitude discrepancies from our results.
We cannot exclude the possibility that thermal chiral EFT is not applicable even at temperatures as low as 100\,MeV,
in which case our thermal mass shift and thermal width for the loosely bound molecule could be much too small.
Another possibility is that the iteration of the integral equations used to obtain self-consistent 
thermal $D^{(*)}$  self energies in Refs.~\cite{Cleven:2019cre,Montana:2022inz} 
give thermal mass shift and thermal width for $X$
that are much too  large at low temperatures.

We have calculated the $D^\ast D$ self energy to NLO in the heavy-meson expansion.
At next-to-next-to-leading order (NNLO) in the heavy-meson expansion,
there will be additional terms that diverge at the  $D^\ast D$ threshold and additional ultraviolet divergences.
It would be interesting to know whether all the NNLO terms that diverge at the $D^\ast D$ threshold
can be absorbed into $D^\ast$ and $D$ propagator corrections
and whether all the ultraviolet divergences at NNLO can be cancelled
by renormalizations of parameters of ZREFT. 
The NLO correction to the binding momentum is suppressed by a factor of $\Delta/M$.
If the NNLO correction to the binding momentum has no such suppression factor, 
it could be comparable to the NLO correction.

We have assumed the interactions of charm mesons and pions in the hadron gas
can be described by $\chi$EFT at leading order at temperatures at least as high as that of kinetic freeze-out.
Some insight into the applicability of thermal $\chi$EFT could be provided by explicit next-to-leading order calculations. 
There have been very few calculations  in thermal $\chi$EFT beyond leading order.
The thermal energy of the pion has been calculated to next-to-leading order by Schenk \cite{Schenk:1993ru}.
Thermal corrections to the pion decay constant and the pion mass
have been calculated to next-to-leading order by Toublan \cite{Toublan:1997rr}.
The corrections are small enough to justify optimism 
towards the applicability of $\chi$EFT at temperatures as high as 115\,MeV.
A next-to-leading order calculation of the thermal energy of a charm meson
should reveal whether thermal $\chi$EFT is applicable at that temperature. 
It could also provide a test of the validity of the self-consistent iteration methods used in  
Refs.~\cite{Cleven:2019cre,Montana:2022inz}.

Our results  suggest that loosely bound charm-meson molecules,
such as  $X(3872)$ and $T_{cc}^+(3875)$, can remain loosely bound and narrow
in the thermal environment of a hadron gas at  sufficiently low  temperature. 
This is consistent with the surprisingly large rate  for the production of $X(3872)$ 
in Pb-Pb collisions at the LHC observed by the CMS collaboration  \cite{CMS:2021znk}.
Our results are more encouraging for the study of loosely bound charm-meson molecules
in heavy-ion collisions than the previous results for $X(3872)$ in  Refs.~\cite{Cleven:2019cre,Montana:2022inz}.
Further studies of $X(3872)$ along with studies of $T_{cc}^+(3875)$  in  heavy-ion collisions
will provide essential insights into the behavior of loosely bound hadronic molecules.

\begin{acknowledgments}
We thank R.~Bruschini for valuable discussions.
This work was supported in part by the U.S.\ Department of Energy under grant DE-SC0011726,
by the National Natural Science Foundation of China (NSFC) under grant 11905112,
by the Alexander von Humboldt Research Foundation,
and by the  NSFC and the Deutsche Forschungsgemeinschaft (DFG)
through the Sino-German Collaborative Research Center TRR110
(NSFC grant 12070131001, DFG Project-ID 196253076-TRR110).
\end{acknowledgments}


\begin{appendix}

\section{Integrals over the Momentum of a Thermal Pion }
\label{app:PionIntegral}

In this Appendix, we give the integrals over the momentum of a pion
that appear in the charm-meson self energies
through next-to-leading order  in the heavy-meson expansion.

\subsection{$\bm{i\, \epsilon}$ Prescriptions}
\label{sec:iepsilon}

Some of the thermal averages over the pion momenta involve  integrals of the form
\beq
\mathcal{F}_n(\sigma) = 
\lim_{\epsilon \to 0^+} \int_0^\infty \mathrm{d}q \, F(q^2)\, \frac{1}{(q^2 - \sigma + i\epsilon)^n} ,
\label{int-n}
\eeq
where $F(q^2)$ is a real-valued function that is smooth as $q^2\to 0$ and decreases rapidly to 0 as $q^2 \to \infty$. 
The real parameter $\sigma$, which can be positive or negative, is small compared to the scale of $q^2$ set by $F(q^2)$.
We would like to expand these functions in powers of $\sigma$.

In the case $n=1$ of Eq.~\eqref{int-n},
the function can be expressed as the sum of a principal-value integral  and the integral of a delta function if $\sigma >0$:
\bqa
\mathcal{F}_1(\sigma)  &=& \int_0^\infty \mathrm{d}q \, F(q^2) 
 \left(\mathcal{P}  \frac{1}{q^2 - \sigma}  - i \pi \, \delta(q^2 - \sigma) \right).
 \nonumber\\
 &=& \int_0^\infty \mathrm{d}q \, \frac{F(q^2) - F(\sigma)}{q^2 - \sigma} 
   - i \, \frac{\pi}{2 \sqrt{\sigma}}\; F(\sigma)  \, \theta(\sigma).
\label{int-1}
\eqa
We have used an identity to express the principal-value integral in terms of an ordinary integral.
The Taylor expansion of the real part of  $\mathcal{F}_1(\sigma)$ can be obtained 
by expanding the integrand in the second line of Eq.~\eqref{int-1} as a Taylor expansion in $\sigma$:
\bqa
\mathrm{Re}\big[\mathcal{F}_1( \sigma)\big] &=& 
 \int_0^\infty \mathrm{d}q \, \frac{F(q^2) - F(0)}{q^2}
+ \sigma \int_0^\infty \mathrm{d}q \, \frac{F(q^2) - F(0) - F^\prime(0)\, q^2 }{q^4}
\nonumber\\
&& + \sigma^2
 \int_0^\infty \mathrm{d}q \, \frac{F(q^2) - F(0) - F^\prime(0) \, q^2 - \tfrac12\, F^{\prime \prime}(0)\,  q^4}{q^6} 
 + \ldots.
\eqa
The first term $F(q^2)/(q^2)^n$ in each integrand can be obtained simply by expanding
the left side of Eq.~\eqref{int-1} in powers of $\sigma$.
The remaining terms in the integrand subtract 
the divergent terms in the Laurent expansion of $F(q^2)/(q^2)^n$ in $q^2$.

The case $n=2$  of Eq.~\eqref{int-n} can be reduced to the case $n=1$ by integrating by parts:
\bqa
\mathcal{F}_2(\sigma)   &=& - \lim_{\epsilon \to 0^+}\int_0^\infty \mathrm{d}q \,  F(q^2) 
\frac{\mathrm{d}\ }{\mathrm{d}q^2} \frac{1}{q^2 - \sigma + i\epsilon} 
 \nonumber\\
 &=&
 \int_0^\infty \mathrm{d}q 
 \left[   F^\prime(q^2)  -  \frac{F(q^2)}{2 q^2} \right]
 \left(\mathcal{P}  \frac{1}{q^2 - \sigma}  - i \pi \, \delta(q^2 - \sigma) \right).
\eqa

\subsection{Integrals over Momentum}
\label{app:IntMom}

The thermal averages  $\mathcal{F}_{cd}$, $\mathcal{G}_{1,cd}$, 
and $\mathcal{G}_{2,cd}$ are defined in Eqs.~\eqref{Fcd}, \eqref{Gcd,n}, and \eqref{Hcd}.
If $\Delta_{cd} > m_{\pi cd}$, their real parts can be expressed in terms of principal-value integrals
 that can be reduced to the form in the first term of the second line of Eq.~\eqref{int-1}:
\begin{subequations}
\bqa
\mathrm{Re}\big[\mathcal{F}_{cd}\big] &=& 
\frac{1}{2\pi^2\, \mathfrak{n}_\pi} \int_0^\infty dq 
\left( \frac{q^4}{\omega_{cdq}} \mathfrak{f}_\pi(\omega_{cdq}) -  \frac{q_{cd}^4}{\Delta_{cd}} \mathfrak{f}_\pi(\Delta_{cd}) \right)  
\frac{1}{q^2 - q_{cd}^2},
\label{ReFcd}
\\
\mathrm{Re}\big[\mathcal{G}_{1,cd}\big] &=& 
\frac{1}{2\pi^2 \mathfrak{n}_\pi} \int_0^\infty dq
\left[  \left( 3q^2 \frac{\omega_{cdq}^2 + \Delta_{cd}^2}{2\omega_{cdq}} 
+ q^4 \frac{\omega_{cdq}^2 - \Delta_{cd}^2}{2 \omega_{cdq}^3} \right) \mathfrak{f}_\pi(\omega_{cdq}) 
- 3  q_{cd}^2 \Delta_{cd} \, \mathfrak{f}_\pi( \Delta_{cd}) 
\right.
\nonumber\\
&& \left. \hspace{3cm}
+ q^4\frac{\omega_{cdq}^2 + \Delta_{cd}^2}{2 \omega_{cdq}^2} \mathfrak{f}^{\,\prime}_\pi(\omega_{cdq}) 
 -  q_{cd}^4\, \mathfrak{f}_\pi^{\, \prime}(\Delta_{cd}) \right]  
\frac{1}{q^2 - q_{cd}^2},
\label{ReGcd,1}
\\
\mathrm{Re}\big[\mathcal{G}_{2,cd}\big] &=& 
\frac{1}{2\pi^2 \mathfrak{n}_\pi} \int_0^\infty dq
\left[  \left( 5 q^4 \frac{\omega_{cdq}^2 + \Delta_{cd}^2}{2 \omega_{cdq}} 
+ q^6 \frac{\omega_{cdq}^2 - \Delta_{cd}^2}{2\omega_{cdq}^3} \right) \mathfrak{f}_\pi(\omega_{cdq}) 
- 5  q_{cd}^4 \Delta_{cd} \, \mathfrak{f}_\pi( \Delta_{cd}) 
\right.
\nonumber\\
&& \left. \hspace{3cm}
+ q^6\frac{\omega_{cdq}^2 + \Delta_{cd}^2}{2\omega_{cdq}^2} \mathfrak{f}^{\,\prime}_\pi(\omega_{cdq}) 
 -  q_{cd}^6\, \mathfrak{f}_\pi^{\, \prime}(\Delta_{cd}) \right]  
\frac{1}{q^2 - q_{cd}^2},
\label{ReGcd,2}
\eqa
\end{subequations}
where $\omega_{cdq}= \sqrt{m_{\pi  cd}^2 + q^2}$ and $q_{cd}^2 = \Delta_{cd}^2 - m_{\pi cd}^2$.
The denominators have a zero at $q_{cd} = \sqrt{\Delta_{cd}^2 - m_{\pi cd}^2}$.
The subtraction of the numerator makes the integral convergent.
If $\Delta_{cd} < m_{\pi cd}$, the subtractions at $\omega_{cdq} =  \Delta_{cd}$ are
not necessary because the integrals are convergent.

The imaginary parts of $\mathcal{F}_{cd}$, $\mathcal{G}_{1,cd}$, and $\mathcal{G}_{2,cd}$ are
nonzero only if $\Delta_{cd} > m_{\pi cd}$.
They can be evaluated analytically using a delta function as in Eq.~\eqref{int-1}:
\begin{subequations}
\bqa
\mathrm{Im}\big[\mathcal{F}_{cd}\big] &=&
- \frac{1}{4\pi \, \mathfrak{n}_\pi}
\left[ \frac{ \mathfrak{f}_\pi(\Delta_{cd})}{\Delta_{cd}} \, q_{cd}^3 \right]
\theta \big(  \Delta_{cd} - m_{\pi cd} \big) ,
\label{ImFcd}
\\
\mathrm{Im}\big[\mathcal{G}_{1,cd}\big] &=& 
- \frac{1}{4\pi\,  \mathfrak{n}_\pi}\,
\big[3  \, \Delta_{cd}\, \mathfrak{f}_\pi(\Delta_{cd}) \,q_{cd} +   \mathfrak{f}_\pi^{\,\prime}(\Delta_{cd}) \, q_{cd}^3 \big] \,
\theta \big(  \Delta_{cd} - m_{\pi cd} \big),
\\
\mathrm{Im}\big[\mathcal{G}_{2,cd}\big] &=&
- \frac{1}{4\pi \, \mathfrak{n}_\pi}
\big[5  \, \Delta_{cd}\, \mathfrak{f}_\pi(\Delta_{cd}) \,q_{cd}^3 +  \mathfrak{f}_\pi^{\,\prime}(\Delta_{cd}) \, q_{cd}^5 \big] \,
\theta \big(  \Delta_{cd} - m_{\pi cd} \big) .
\label{ImGcd,2}
\eqa
\end{subequations}

\subsection{Expansions in Isospin Splittings}
\label{sec:ExpandSplitting}

The thermal average over the Bose-Einstein distribution for a pion is defined in Eq.~\eqref{<F>}.
The thermal averages  $\mathcal{F}_{cd}$, $\mathcal{G}_{1,cd}$, 
and $\mathcal{G}_{2,cd}$ defined in Eqs.~\eqref{Fcd}, \eqref{Gcd,n}, and \eqref{Hcd} 
depend on $\Delta_{cd}^2- m_{\pi cd}^2$, which is linear in isospin splittings.
These thermal averages can be expanded in powers of $\Delta_{cd}^2 - m_{\pi cd}^2$
using the results presented in Section~\ref{sec:iepsilon}.
Their real parts can be expanded in integer powers of isospin splittings divided by $m_\pi$.
The leading terms in the expansions of the real parts of $\mathcal{F}_{cd}$, 
$\mathcal{G}_{1,cd}$, and $\mathcal{G}_{2,cd}$ are
\begin{subequations}
\bqa
\mathrm{Re}\big[\mathcal{F}_{cd}\big] &\approx& 
\left \langle \frac{1}{\omega_q} \right\rangle ,
\label{ReFcd:expand}
\\
\mathrm{Re}\big[\mathcal{G}_{1,cd}\big] &\approx& 
\left \langle \frac{\omega_q^2+ m_\pi^2}{q^2\, \omega_q} \right\rangle 
,
\label{ReGcd,1:expand}
\\
\mathrm{Re}\big[\mathcal{G}_{2,cd}\big] &\approx& 
\left\langle \frac{\omega_q^2 + m_\pi^2}{\omega_q} \right\rangle  .
\label{ReGcd,2:expand}
\eqa
\end{subequations}
The imaginary parts of $\mathcal{F}_{cd}$, $\mathcal{G}_{1,cd}$, and $\mathcal{G}_{2,cd}$ are
nonzero only if $\Delta_{cd} > m_{\pi cd}$.
They can be expanded in half-integer powers of isospin splittings divided by $m_\pi$.
The leading terms in the expansions of their imaginary parts  are
\begin{subequations}
\bqa
\mathrm{Im}\big[\mathcal{F}_{cd}\big] &\approx&
\frac{\mathfrak{f}_\pi(m_\pi)}{4\pi\, \mathfrak{n}_\pi} 
\left( -\frac{1}{m_\pi} \, q_{cd}^3 \right)
\theta \big(  \Delta_{cd} - m_{\pi cd} \big) ,
\label{ImFcdleading}
\\
\mathrm{Im}\big[\mathcal{G}_{1,cd}\big] &\approx& 
\frac{\mathfrak{f}_\pi(m_\pi)}{4\pi\,\mathfrak{n}_\pi} \,\big( - 3\, m_\pi\,  q_{cd} \big)\,
\theta \big(  \Delta_{cd} - m_{\pi cd} \big),
\label{ImGcd1leading}
\\
\mathrm{Im}\big[\mathcal{G}_{2,cd}\big] &\approx&
\frac{\mathfrak{f}_\pi(m_\pi)}{4\pi\,\mathfrak{n}_\pi} \, \big( - 5\, m_\pi \,  \, q_{cd}^3 \big)\,
\theta \big(  \Delta_{cd} - m_{\pi cd} \big) .
\label{ImGcd2leading}
\eqa
\end{subequations}

\subsection{Thermal Average $\bm{\mathcal{H}_{cd}}$}
\label{sec:Hcd}

As $\epsilon \to 0^+$, the  thermal average $\mathcal{H}_{cd}$ defined in Eq.~\eqref{Hcd} 
has a contribution that  diverges as $1/\epsilon$ if  $\Delta_{cd} > m_{\pi cd}$.
 If we take into account the decay widths in the charm-meson propagators, 
 this thermal average is replaced by
\bqa
\mathcal{H}_{cd} &=& \frac14 \sum_\pm
\left\langle  \frac{q^2}{\omega_{cdq}\, \big[ \pm\omega_{cdq} - \Delta_{cd} + i (\Gamma_{\ast c}-\Gamma_d)/2\big]}   
\right.
\nonumber\\
&&\hspace{2cm}\left. \times
\left( \frac{1}{\pm\omega_{cdq} - \Delta_{cd} + i \Gamma_{\ast c}}
+ \frac{1}{\pm\omega_{cdq} - \Delta_{cd} - i \Gamma_d} \right)
 \right\rangle.
 \label{Hab-Gamma}
\eqa
The thermal average can be expanded in powers of isospin splittings.
The leading  term in the expansion is
\beq
\mathcal{H}_{cd} \approx
\left \langle \frac{\omega_q^2+ m_\pi^2}{q^2\, \omega_q} \right\rangle.
\eeq
If the thermal average in Eq.~\eqref{Hab-Gamma} is simplified by setting $\Gamma_{\ast c} =\Gamma_d =\Gamma$,
$\mathcal{H}_{cd}$ becomes real valued.
If $ \Delta_{cd} > m_{\pi cd}$,
$\mathcal{H}_{cd}$ has a term proportional to  $q_{cd}^3$ that
diverges as $1/\Gamma$ as $\Gamma \to 0$:
\beq
\mathcal{H}_{cd} \longrightarrow 
\frac{\mathfrak{f}_\pi( \Delta_{cd})}{4 \pi \,\mathfrak{n}_\pi}\,  \frac{1}{\Gamma}\, q_{cd}^3\,
\theta\big( \Delta_{cd} - m_{\pi cd} \big).
\eeq
If $\Gamma_{\ast c} \neq \Gamma_d$, $\mathcal{H}_{cd}$ also has  imaginary terms 
that diverge as  $\Gamma_{\ast c} \to 0$ and $\Gamma_d \to 0$.

\section{Integrals over the Relative Momentum of a Charm-meson Pair}
\label{app:MomIntegral}

In this Appendix, we give the integrals over the relative momentum of a charm-meson pair
that appear in the $D^\ast D$ self energy through NLO in the heavy-meson expansion.
The only integral in the $D^\ast D$ self energy at leading order is
\beq
 \int \!\!\frac{d^3k}{(2\pi)^3}  \, 
 \frac{1}{\big[ E - (\varepsilon_\ast + \varepsilon) - \bm{k}^2/(2M)- (\bm{P}-\bm{k})^2/(2M_\ast)  + i \epsilon \big]^2}
= \frac{\mu^2}{2 \pi} \, \frac{1}{S_0(E_\mathrm{cm}) },
\label{softint}
\eeq
where $\mu = M_\ast M/(M_\ast + M)$ and
$S( E_\mathrm{cm} )$ is the function of the center-of-mass energy in Eq.~\eqref{S0-Ecm}.
The additional integrals that appear in the $D^\ast D$ self energy at  NLO are
\begin{subequations}
\bea
\int \!\!\frac{d^3k}{(2\pi)^3}  \, 
 \frac{1}{ E - (\varepsilon_\ast + \varepsilon) - \bm{k}^2/(2M)- (\bm{P}-\bm{k})^2/(2M_\ast)  + i \epsilon }
=&&\, - \frac{\mu}{2 \pi} \, \big[  \Lambda - S_0(E_\mathrm{cm})  \big],
\label{int1}
\\
 \int \!\!\frac{d^3k}{(2\pi)^3}  \, 
 \frac{\bm{k}^2}{ \big[E - (\varepsilon_\ast + \varepsilon) - \bm{k}^2/(2M)- (\bm{P}-\bm{k})^2/(2M_\ast)  + i \epsilon\big]^2 }
\nonumber\\
&& \hspace{-3.5cm}
=\frac{\mu^2}{\pi}  \left[  \Lambda - \frac32 S_0(E_\mathrm{cm})  
+\frac{M^2}{2 M_X^2}  \frac{P^2 }{S_0(E_\mathrm{cm})}\right],
\label{int2-k}
\\
\int \!\!\frac{d^3k}{(2\pi)^3}  \, 
\frac{(\bm{P}-\bm{k})^2}{ \big[E - (\varepsilon_\ast + \varepsilon) - \bm{k}^2/(2M)- (\bm{P}-\bm{k})^2/(2M_\ast)  + i \epsilon\big]^2 }
\nonumber\\
&&  \hspace{-3.5cm}
=\frac{\mu^2}{\pi}  \left[ \Lambda - \frac32 S_0(E_\mathrm{cm})  
+ \frac{M_\ast^2}{2 M_X^2}  \frac{P^2 }{S_0(E_\mathrm{cm})} \right],
\label{int2-Pmk}
\eea
\end{subequations}
where $M_X = M_\ast + M$.
We have imposed  a large momentum cutoff $|\bm{k}| < (\pi/2)\Lambda$ to regularize the ultraviolet divergences.

\section{Feynman Rules for Charm Mesons and Pions}
\label{app:Feynman}

In this appendix, we give the Feynman rules for charm mesons and pions used in this paper.
We use the latin letters $i, j$ for pion flavors, $a, b$
for charm-meson flavors, and $m, n, k$ for Cartesian indices.

\subsection{Propagators}
The propagator for a relativistic pion
with 4-momentum $q$ is
\begin{equation}
    \frac{i \delta^{ij}}{q^2 - m_{\pi i}^2 + i \epsilon},
\end{equation}
where $m_{\pi i}$ is the mass of the pion with flavor $i$.
The propagator for a nonrelativistic pseudoscalar charm meson $D$ with 
energy $E$ (relative to the kinetic mass $M$)
and 3-momentum $\bm p$ is
\begin{equation}
    \frac{i \delta_{ab}}{E - \varepsilon_a - \bm p^2/(2M) + i \epsilon},
\end{equation}
where $\varepsilon_a = M_a - M$ is the rest energy of $D^a$.
The propagator for a nonrelativistic vector charm meson $D^\ast$ with
energy $E$ (relative to the kinetic mass $M_\ast$)
and 3-momentum $\bm p$ is 
\begin{equation}
    \frac{i \delta_{ab}\; \delta^{mn}}
    {E - \varepsilon_{\ast a} - \bm p^2/ (2M_\ast) + i\epsilon},
\end{equation}
where $\varepsilon_{\ast a} = M_{\ast a} - M_\ast$ is the rest energy of $D^{\ast a}$.
The $D^{\ast}$ decay width can be taken into account by replacing $i \epsilon$ with
$i\Gamma_{\ast a} / 2$.

\subsection{Vertices}
In heavy-hadron $\chi$EFT at leading order,
the vertices for the interactions between charm mesons and pions are determined by 
the pion decay constant $f_\pi$ and a dimensionless constant $g_\pi$.
The $D^{(\ast)}\pi \to D^{(\ast)} \pi$ contact vertices, 
with the incoming and outgoing pions having 3-momenta $\bm{q}$ and $\bm{q}^\prime$, are
\begin{subequations}
    \begin{align}
        D^{ a}\pi^i (\bm{q})\to D^{ b} \pi^j(\bm{q}^\prime)\!\!
        &:\: +\frac{i}{4 f_\pi^2} [\sigma^i, \sigma^j]_{ab} (\omega_q + \omega_{q^\prime}),
        \\
        D^{\ast a}_m \pi^i(\bm{q}) \to D^{\ast b}_n \pi^j(\bm{q}^\prime)\!\!
        &:\: +\frac{i}{4 f_\pi^2} [\sigma^i, \sigma^j]_{ab} \; \delta^{mn} (\omega_q + \omega_{q^\prime}),
    \end{align}
\end{subequations}
where $\sigma^i$ is a Pauli matrix and $\omega_q = \sqrt{m_\pi^2 + q^2}$.
The vertices for the transitions $D^{(\ast)} \to D^{(\ast)} \pi$,  with the outgoing pion having  3-momentum $\bm q$, are
\begin{subequations}
\begin{eqnarray}
         D^a \to D^{\ast b}_n \pi^i(\bm{q})\!:~
        &&+ i\frac{g_\pi}{\sqrt{2}f_\pi} \sigma_{ab}^i\; q^n,
        \\
         D^{\ast a}_m \to D^b \pi^i(\bm{q})\!:~
        && - i\frac{g_\pi}{\sqrt{2}f_\pi} \sigma_{ab}^i\; q^m,
        \\
         D^{\ast a}_m \to D^{\ast b}_n \pi^i(\bm{q})\!:~
        &&+  i\frac{g_\pi}{\sqrt{2}f_\pi} \sigma_{ab}^i\; \varepsilon^{mnk} q^k,
        \label{D*toD*pi}
\end{eqnarray}
\end{subequations}
where $\varepsilon_{mnk}$ is the Levi-Civita symbol.
The corresponding vertices for the transitions $D^{(\ast)} \pi \to D^{(\ast)}$,
with the  incoming pion having  3-momentum $\bm q$, are obtained 
by replacing $\bm q$ with $-\bm q$.

\end{appendix}


\end{document}